\documentclass[lettersize,journal]{IEEEtran}
\newtheorem{Remark}{\it Remark}[section]

\newtheorem{Proposition}{\it Proposition}[section]
\newtheorem{Lemma}{\it Lemma}[section]

\newcommand*{\QEDA}{\hfill\ensuremath{\blacksquare}} 
 
\makeatletter
\newcommand{\Rmnum}[1]{\expandafter\@slowromancap\romannumeral #1@}
\makeatother
\usepackage{tabu}
\usepackage{verbatim}
\usepackage{graphicx}
\usepackage{bm}
\usepackage{multicol}
\usepackage{setspace}
\usepackage{subfigure}
\usepackage{float}
\usepackage{amsmath}
\usepackage{epstopdf}
\allowdisplaybreaks[4]
\usepackage{fancyhdr} 
\usepackage{indentfirst}
\usepackage{enumerate}
\usepackage{lipsum}
\usepackage{amssymb}
\usepackage{stfloats}
\usepackage{bm}
\usepackage{threeparttable}
\usepackage{CJK}
\usepackage[justification=centering]{caption}
\usepackage{makecell}
\usepackage{caption}
\usepackage{cases}
\usepackage{algorithm}
\usepackage{algpseudocode}
\usepackage{graphics}

\usepackage{color}
\usepackage{epsfig}
\usepackage{cite}

\def\BibTeX{{\rm B\kern-.05em{\sc i\kern-.025em b}\kern-.08em
		T\kern-.1667em\lower.7ex\hbox{E}\kern-.125emX}}

\begin{document}
	\title{Performance Trade-off of Integrated Sensing and Communications for Multi-User Backscatter Systems}
    \author{Yuanming Tian, Dan Wang, Chuan~Huang,~\IEEEmembership{Member, IEEE}, and Wei~Zhang,~\IEEEmembership{Fellow, IEEE} 
	
	\thanks{
	    This article will be presented in part at IEEE International Symposium on Personal, Indoor and Mobile Radio Communications Workshops (PIMRC Workshops), Valencia, Spain, Sep. 2024\cite{PIMRC}.
		
		Yuanming Tian is with the Shenzhen Future Network of Intelligence Institute, the School of Science and Engineering, and the Guangdong Provincial Key Laboratory of Future Networks of Intelligence, The Chinese University of Hong Kong (Shenzhen), Shenzhen 518172, China (e-mail: yuanmingtian@link.cuhk.edu.cn).
		
		Dan Wang is with the Department of Broadband Communications, Peng Cheng Laboratory, Shenzhen 518055, China (e-mail: wangd01@pcl.ac.cn).	
		
		Chuan Huang is with the School of Science and Engineering, the Shenzhen Future Network of Intelligence Institute, and the Guangdong Provincial Key Laboratory of Future Networks of Intelligence, The Chinese University of Hong Kong (Shenzhen), Shenzhen 518172, China (e-mail: huangchuan@cuhk.edu.cn).
		
		Wei Zhang is with the School of Electrical Engineering and Telecommunications, The University of New South Wales, Sydney, NSW 2052, Australia
		(e-mail: w.zhang@unsw.edu.au).
	}
		
	}

\maketitle

	\begin{abstract}
    This paper studies the performance trade-off in a multi-user backscatter communication (BackCom) system for integrated sensing and communications (ISAC), where the multi-antenna ISAC transmitter sends excitation signals to power multiple single-antenna passive backscatter devices (BD), and the multi-antenna ISAC receiver performs joint sensing (localization) and communication tasks based on the backscattered signals from all BDs. Specifically, the localization performance is measured by the Cram\'{e}r-Rao bound (CRB) on the transmission delay and direction of arrival (DoA) of the backscattered signals, whose closed-form expression is obtained by deriving the corresponding Fisher information matrix (FIM), and the communication performance is characterized by the sum transmission rate of all BDs. Then, to characterize the trade-off between the localization and communication performances, the CRB minimization problem with the communication rate constraint is formulated, and is shown to be non-convex in general. By exploiting the hidden convexity, we propose an approach that combines fractional programming (FP) and Schur complement techniques to transform the original problem into an equivalent convex form. Finally, numerical results reveal the trade-off between the CRB and sum transmission rate achieved by our proposed method.
	\end{abstract}

		
	\section{Introduction}
	
	Backscatter communication (BackCom) has recently drawn substantial attentions from both the academic and industrial spheres, and is regarded as one promising solution for low-power communications for the next generation mobile communication networks \cite{FutureIoT,6G,Low,AddOne,AddThree}. In BackCom systems, passive backscatter device (BD), e.g., radio frequency identification (RFID), can harvest energy from an external radio frequency (RF) source to power its circuit, and then modulates its own information onto the external RF signal to accomplish its transmissions \cite{Air}. In applications, BDs in various BackCom systems are powered by different RF sources, i.e., signals from mono-static readers in mono-static BackCom (MoBack) systems\cite{MoBack}, dedicated RF excitation signals in bi-static BackCom (BiBack) systems\cite{BiBack}, and ambient RF signals (e.g., those from television towers and cellular base stations (BSs)) in ambient BackCom (AmBack) systems \cite{AmBack}. Without carrying bulky batteries or being powered by power gird, BackCom exhibits significant advantages over conventional wireless RF communication systems, i.e., low cost, extremely low energy consumption \cite{MASK}, flexible deployment, and effortless maintenance \cite{Advantage2}, making them suitable for a variety of applications, such as industrial automation, precision agriculture, and smart cities \cite{App}.  
	
	Integrated sensing and communications (ISAC) is another emerging technology for the next generation mobile communication networks, which achieves seamless integration of sensing and wireless communications by sharing hardware, spectrum, and energy \cite{ISAC1,ISAC2,ISAC3}. In such dual-functional systems, sensing function aims to collect and analyze echo signals from the surrounding environment, thereby facilitating the extraction of valuable information about the targets \cite{ISAC1}, e.g., distance, angle, and radial velocity. On the other hand, communication function mainly focuses on the information transmissions between the BS and users via specifically tailored signals \cite{ISAC2}. The deep integration of these two functions is expected to significantly improve the utilization efficiency of scarce spectrum and energy resources, while reducing both the hardware and signaling costs\cite{ISAC3}. These advantages endow ISAC with immense potential for the various applications, such as vehicle-to-everything (V2X) communications, environmental monitoring, and smart home \cite{ISAC1,ISAC2,ISAC3}.
    
    Building upon the above two concepts, ISAC technology can be effectively merged with BackCom to extend its functionality and application ranges \cite{New}. The fusion of these two technologies, referred to as integrated sensing and backscatter communications (ISABC) \cite{ISABC} or BackCom for ISAC \cite{ISAC-AmBC}, enables passive BDs to serve as sensors, providing valuable environmental information by modulating information onto the excitation signal and then reflecting the modulated signals to the receivers. BackCom for ISAC, facilitating a wide range of applications, such as end-to-end vehicular safety-awareness system \cite{Future} and intelligent transportation and delivery\cite{Advantage}, has demonstrated numerous advantages, e.g., enhanced integration level \cite{ISABC} and improved energy efficiency \cite{Advantage}, and has been widely investigated in \cite{ISAC-Loc,ISAC-radar,ISABC,ISAC-AmBC}. Specifically, the authors in \cite{ISAC-Loc} considered an AmBack system for ISAC, consisting of multiple full-duplex (FD) BSs, multiple passive BDs, and multiple backscatter receivers, and utilized a combination of the generalized Prony with homologous matching (GPHM) method and multi-source constant modulus algorithm (MSCMA) to simultaneously accomplish the localization task and symbiotic communication. The authors in \cite{ISAC-radar} leveraged the radar clutter as an excitation signal to power passive BDs, and accordingly proposed two distinct encoding/decoding strategies, namely frame-by-frame encoding and frame-differential encoding, to enhance the communication performance of the considered system. The authors in \cite{ISABC} considered an AmBack system for ISAC, consisting of one FD BS, one passive BD, and one cooperative user, and derived the closed-form expressions of the communication rates for both the user and BD, as well as the sensing rate at the BS. By considering multiple BD, the authors in \cite{ISAC-AmBC} designed a two-phase link-layer protocol, which includes one training phase and one localization and information transmission phase, to accomplish joint detection and localization of all BDs.
    
    In those prior works on ISAC systems and BackCom systems for ISAC, sensing rate \cite{ISABC}, transmit beampattern \cite{Beampattern}, and recognition accuracy \cite{Recognition} have been adopted as the performance metrics for sensing tasks. In general, the Cram\'{e}r-Rao bound (CRB), which serves as the theoretical lower limit on the variance of any unbiased estimator \cite{Steven}, is widely adopted as an analytically tractable metric for assessing the performance of general sensing tasks \cite{Time,CRB-single,CRB-multiple}. Specifically, the authors in \cite{Time} investigated one single-target localization problem with the help of intelligent reflecting surface (IRS) in a sensing-only system, where the closed-form expression for the CRB on the transmission delay and direction of arrival (DoA) was derived\footnote{In this work, localization was accomplished by determining the direction of the target with respect to the IRS and the distance between them via the DoA and transmission delay, respectively. Additionally, radar signal properties were utilized to derive the CRB on these parameters, which generally lacks applicability to ISAC systems.}. The authors in \cite{CRB-single} studied a multi-user ISAC system over a broadcast channel with one single target, and characterized the CRB on the DoA for the single target in two specific scenarios, namely the point and extended target models, respectively. Furthermore, the authors in \cite{CRB-multiple} studied a multi-user ISAC system over a multicast channel with multiple targets, in which a matrix-form expression for the CRB on the reflection coefficient and DoA for all targets was derived. Nevertheless, to our best knowledge, how to derive a closed-from expression for the CRB on the transmission delay and DoA for multiple targets under the BackCom for ISAC scenario remains an unexplored problem.
    
    To address this challenge, this paper studies a multi-user BackCom system for ISAC, consisting of one multi-antenna ISAC transmitter, multiple single-antenna passive BDs, and one multi-antenna ISAC receiver. Under this scenario, the ISAC transmitter sends excitation signals to power all passive BDs, and the ISAC receiver performs joint sensing (localization) and communication tasks based on the backscattered signals from all BDs. However, there still exists a trade-off between localization and communication tasks, due to they share wireless resources within one integrated hardware system. As a result, characterizing this trade-off is essential for the considered system. Therefore, we characterize the performance bound on the localization task and propose an analytical framework for the multi-user BackCom system for ISAC, focusing on optimal trade-off achievement through the sample covariance matrix design. The main contributions of this paper are summarized as follows:
    \vspace{-1pt}
    \begin{itemize} 
    \item First, we derive the transmitted signal model at the ISAC transmitter, the signal model at BDs, and the received signal model at the ISAC receiver in a sequential manner. Drawing upon the received signal model, we derive the Fisher information matrix (FIM) corresponding to the transmission delay and DoA of the backscattered signals, and accordingly obtain the closed-form expression of the CRB on these parameters.

    \item Then, we formulate an optimization problem that aims to minimize the obtained CRB by designing the sample covariance matrix at the ISAC transmitter, subject to the constraints of the sum transmission rate and power budget. Unfortunately, the formulated CRB minimization problem exhibits non-convexity and is thus difficult to be solved. To address this issue, we propose an approach that combines fractional programming (FP) and Schur complement techniques, which equivalently transforms the original non-convex problem into a convex one. 

    \end{itemize}
    
	The remainder of this paper is organized as follows. Section \ref{System Model} introduces the system model of the multi-user BackCom system for ISAC. Section \ref{Performance} conducts an analysis for the sensing and communication performances of the considered system, and derives the closed-from expression on the CRB for parameter estimations and sum transmission rate of all BDs. Section \ref{Optimization} presents the transmit beamforming design for CRB minimization under the constraints of the sum transmission rate and power budget. Section \ref{Simulation} presents the numerical results. Finally, Section \ref{conclusion} summarizes this paper.
	
	\emph{Notations}: We use lowercase letters, e.g., $x$, boldface lowercase letters, e.g., $\bm{x}$, and boldface uppercase letters, e.g., $\bm{X}$, to denote scalars, vectors, and matrices, respectively. $\mathbb{C}$, $\mathbb{R}$, and $\mathbb{R}^{+}$ denote the sets of all complex, real, and positive real numbers, respectively. $\|\bm{x}\|$ denotes the Euclidean norm of vector $\bm{x}$. $\bm{X}^T$ and $\bm{X}^H$ denote the transpose and conjugate transpose of matrix $\bm{X}$, respectively. $\mbox{vec}(\bm{X})$ represents stacking the columns of matrix $\bm{X}$ into a column vector. $\mbox{Tr}(\bm{S})$ and $\bm{S}^{-1}$ denote the trace and inverse of square matrix $\bm{S}$, respectively. $\text{diag}\left\{x_1,x_2,\cdots,x_N\right\}$ denotes the diagonal matrix, with $x_1,x_2,\cdots,x_N$ being the diagonal elements. $\bm{I}_{M}$
	denotes the identity matrix of size $M\times M$.
	The distribution of a circularly symmetric complex Gaussian (CSCG) random vector with mean vector $\bm{\mu}$ and covariance matrix $\bm{\Sigma}$ is denoted by $\mathcal{CN}(\bm{\mu},\bm{\Sigma})$. $\mathcal{R}(x)$ denotes the real part of complex number $x$. $j=\sqrt{-1}$ denotes the imaginary unit. $\mathbb{E}\left\{\cdot\right\}$ denotes the statistical expectation. log($\cdot$) and ln($\cdot$) denote the logarithm functions with bases 2 and e, respectively. $[L]$ denotes the set of all positive integers {no bigger than $L$.
	\vspace{4pt}
	\section{System Model}
	\label{System Model}	
    \subsection{System Overview}

    As depicted in Fig.\ref{Backscatter System}, we consider a multi-user BackCom system for ISAC, consisting of one ISAC transmitter, multiple BDs, and one ISAC receiver. In particular, this system accommodates $L$ BDs, $L\geq1$, denoted by a set $\mathcal{L}=\left\{1,\ldots,L\right\}$, and each BD is a passive device that can harvest energy from the excitation signal to accomplish information transmissions. The ISAC transmitter and the ISAC receiver are equipped with $M_t$ and $M_r$ antennas, respectively, and the BD is equipped with one single antenna. Without loss of generality, considering a two-dimensional (2D) Cartesian coordinate system, the locations of the ISAC transmitter, the ISAC receiver, and the $l$-th BD, $\forall l \in \mathcal{L}$, are denoted as $(x_T,y_T)$, $(x_R,y_R)$, and $(x_l,y_l)$, respectively. Thus, the Euclidean distances between the ISAC transmitter and the $l$-th BD, the ISAC receiver and the $l$-th BD, and the ISAC transmitter and the ISAC receiver are given as
    $d_{T,l}=\sqrt{(x_T-x_l)^2+(y_T-y_l)^2}$,  $d_{R,l}=\sqrt{(x_R-x_l)^2+(y_R-y_l)^2}$, and $d_{0}=\sqrt{(x_T-x_R)^2+(y_T-y_R)^2}$, respectively.

    \begin{figure}[t]
	\centering
	\includegraphics[scale=0.58]{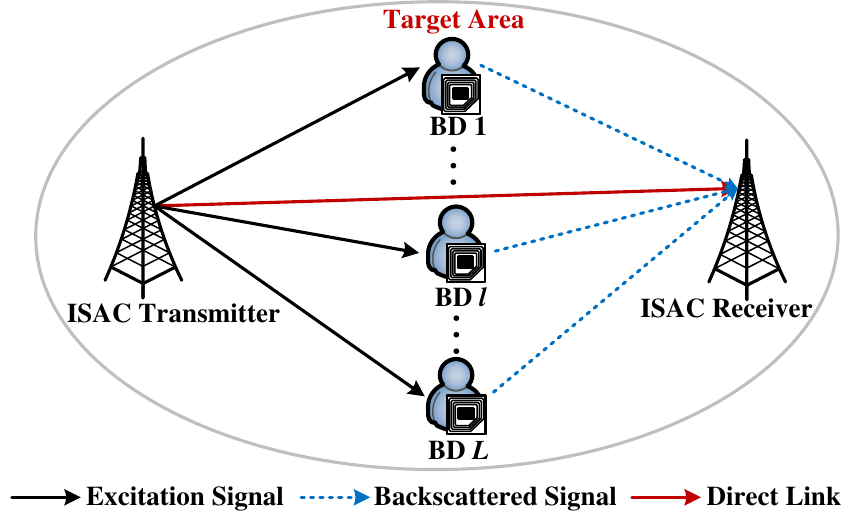}
	\caption{System model of multi-user BackCom system for ISAC.}
	\label{Backscatter System}
    \end{figure} 
    
	The signal transmission and processing process of the considered multi-user BackCom system for ISAC is succinctly outlined as follows: First, the ISAC transmitter sends an excitation signal to all BDs in the designated target area; then, each passive BD, upon harvesting an ample quantity of energy, proceeds to modulate its individual information symbols onto the excitation signals, and reflects the modulated signals towards the ISAC receiver; finally, the ISAC receiver receives the backscattered signals from all BDs, and discretizes the received signals with a fixed  sampling interval.
     
    \subsection{Transmitted Signal Model}
            \begin{figure}[htb]
    	\centering
    	\includegraphics[scale=0.62]{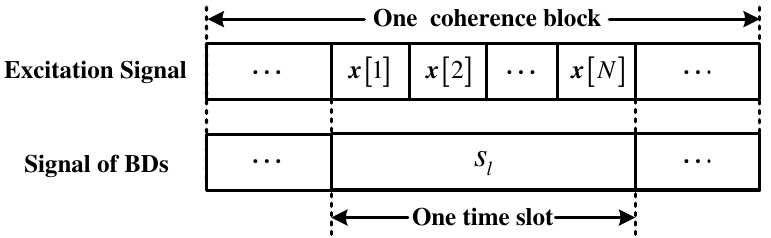}
    	\caption{Frame structure of multi-user BackCom signals.}
    	\label{Frame}
    \end{figure}        
    For simplicity, we consider the ISAC transmission over one coherence block \cite{block} consisting of $T$ time slots, where the duration of each time slot is denoted as $\Delta T$. As depicted in Fig. \ref{Frame}, during each time slot, the ISAC transmitter sends $N$ excitation symbols, each of which is with duration $\Delta t$. Based on the above setup, the excitation signal sent from the ISAC transmitter within the considered time slot is expressed as
     \begin{align}
    	\label{transmit}
    	\bm{x}(t)=\sum_{n=1}^{N} \bm{x}[n] g\big(t-(n-1)\Delta t\big),
    \end{align} 
    where $\bm{x}[n]\in\mathbb{C}^{M_t\times1}$, $n\in\left\{1,\ldots,N\right\}$, is the $n$-th symbol of the excitation signal, and is independent and identically distributed (i.i.d.) across different $n$; $g(t)$ is the transmit pulse function with domain $[0,\Delta t]$. Hence, the transmission power at the ISAC transmitter is given as   
    \begin{equation}
    \label{power}
    \mathbb{E}\left\{\|\bm{x}(t)\|^2\right\}=p_{g}\mbox{Tr}\left(\bm{R}_x\right)\leq P_0,
    \end{equation}
    where $p_{g}$ is the average power of the transmit pulse function $g(t)$ and is set as a unit value for simplicity \cite{Waveform}, i.e., $p_{g}=\frac{1}{\Delta t}\int_{0}^{\Delta t} |g(t)|^2 dt=1$; $P_0$ is the power budget within the considered time slot; $\bm{R}_x$ is the sample covariance matrix of the excitation signal, corresponding to the transmit beamforming vectors \cite{TransmitBeamforming} to be designed, i.e.,
    \begin{equation}
    	\bm{R}_x=\frac{1}{N}\sum_{n=1}^{N}\bm{x}[n]\bm{x}[n]^H.
    \end{equation}
 
    \subsection{Signal Model at BDs}
    By transmitting the excitation signal $\bm{x}(t)$ in \eqref{transmit} to all BDs, the received signal at the $l$-th BD is given as
    \begin{equation}
    y_{l}(t)= \sqrt{\eta(d_{T,l})}\bm{a}_t^{T}(\theta_l)\bm{x}(t-\tau_{T,l}),
    \end{equation}
    where $\eta(d_{T,l})$ is the large-scale fading coefficient of the link from the ISAC transmitter to the $l$-th BD and is modeled as $\eta(d)=\zeta\cdot d^{-\gamma}$ \cite{LargeScale}, with $\zeta$, $d$, and $\gamma$ being the path loss at the reference distance of $1$ meter (m), the propagation distance, and the path loss exponent, respectively; $\bm{a}_{t}(\theta_l)\in\mathbb{C}^{M_t\times 1}$ denotes the transmit steering vector, with $\theta_{l}$ being the direction of departure (DoD) at the ISAC transmitter with respect to the $l$-th BD; $\tau_{T,l} = d_{T,l}/c$ is the propagation delay from the ISAC transmitter to the $l$-th BD, with $c$ being the speed of light in free space. Note that the noise at all BDs is disregarded, as demonstrated in \cite{BiBack,AmBack}, and \cite{Noise}, due to the fact that the BDs in the backscatter system only consist of passive components. Here, we consider a typical scenario where uniform linear arrays (ULAs) are deployed at the ISAC transmitter, and thus the transmit steering vector is given as \cite{TransmitBeamforming}
     \begin{equation}
    	\bm{a}_{t}(\theta_l)= \left[1,e^{j2\pi\frac{d_T}{\lambda}\sin(\theta_l)},\cdots,e^{j2\pi\frac{(M_t-1)d_T}{\lambda}\sin(\theta_l)}\right]^T,
    \end{equation}
    where $d_T$ is the distance between any two adjacent antennas and $\lambda$ is the carrier wavelength.
    
    In comparison to the conventional ISAC transmitter \cite{Transmitter}, BD features a much simpler hardware design and operates with ultra-low power consumption \cite{6G,MASK}. Consequently, the modulation bandwidth of the BD is significantly smaller than that of the transmitter, and thus in this paper we consider the scenario that the BDs transmit one symbol during the whole time slot, as shown in Fig. \ref{Frame}. Moreover, in the considered BackCom system, modulation at each BD is implemented by varying the amplitude of the excitation signal \cite{MASK}, and the transmitted information symbol of the $l$-th BD within the considered time slot is denoted as $s_l\in\mathbb{R}^{+}$, which is generated from the modulation alphabet set $\mathcal{S}_{D}$, i.e., $s_{l}\in \mathcal{S}_D=\left\{s_1,s_2,\cdots,s_{D}\right\}$,
    and each element in $\mathcal{S}_{D}$ falls within the range $[0,1]$. Based on the above setup, the backscattered signal sent from the $l$-th BD is expressed as 
    \begin{equation}
    \label{backscattered}
     r_l(t)=s_l\sqrt{\varrho\cdot \eta(d_{T,l})}\bm{a}_t^T(\theta_l)\bm{x}(t-\tau_{T,l}-\tau_0),
    \end{equation}
     where $\varrho\in [0,1]$ is reflection coefficient, i.e., the fraction of the power backscattered by each BD, and $\tau_0$ is the constant response delay at each BD.
    
    \vspace{-5pt}
    \subsection{Received Signal Model}
    In this paper, we consider the scenario that excitation signal $\bm{x}(t)$ is perfectly known to the ISAC receiver, and the direct link interference, i.e., the excitation signal transmitted directly from the ISAC transmitter to the ISAC receiver, can be effectively eliminated at the receiver, as demonstrated in \cite{Interference} and \cite{neglect}. Furthermore, it is assumed that the interactions among BDs is neglected, which has been widely adopted in the related works\cite{neglect},\cite{neglect2}. Therefore, the signal received by the ISAC receiver is expressed as the superposition of the receiver noise, the clutter reflected from nearby obstacles, and the backscattered signals from all BDs, i.e., 
    \begin{equation}
    \begin{aligned}
    \label{sensing}
    \bm{y}(t)=& \sum_{l=1}^{L} s_l\sqrt{\varrho\cdot\eta(d_{T,l})\eta(d_{R,l})}\bm{a}_{r}(\phi_{l})\bm{a}_{t}^T(\theta_l)\\
    &\qquad\qquad\;\times\bm{x}(t-\tau_{T,l}-\tau_{0}-\tau_{R,l})+\bm{c}(t) +\bm{z}(t),\\
    =&\sum_{l=1}^{L}s_l\alpha_l\bm{H}_{l}\bm{x}(t-\tau_{0,l})+\bm{c}(t) +\bm{z}(t),
    \end{aligned}
    \end{equation} 
    where $\eta(d_{R,l})$ is the large-scale fading coefficient of the link from the $l$-th BD to the ISAC receiver; $\bm{a}_r(\phi_{l})\in\mathbb{C}^{M_r\times 1}$ is the receive steering vector, with $\phi_{l}$ being the DoA at the ISAC receiver with respect to the $l$-th BD; $\tau_{R,l} = d_{R,l}/c$ is the propagation delay from the $l$-th BD to the ISAC receiver; for notation convenience, we denote $\sqrt{\varrho\cdot\eta(d_{T,l})\eta(d_{R,l})}$ and $\bm{a}_{r}(\phi_{l})\bm{a}_{t}^T(\theta_{l})$ as $\alpha_l$ and $\bm{H}_l$, respectively; $\tau_{0,l}=\tau_{T,l}+\tau_{0}+\tau_{R,l}$ is the total transmission delay from the ISAC transmitter to the ISAC receiver via the $l$-th BD; $\bm{c}(t)\sim\mathcal{CN}(0,\sigma_{c}^2\bm{I}_{M_r})$ is the clutter, including the reflected signals from the obstacles within the designated target area \cite{clutter}, with $\sigma_{c}^2$ being the average clutter power; and $\bm{z}(t)\sim\mathcal{CN}(0,\sigma_{z}^2\bm{I}_{M_r})$ is the CSCG noise, with $\sigma_{z}^2$ being the average noise power. Here, similar to the ISAC transmitter, ULAs are deployed at the ISAC receiver, and thus the receive steering vector is given as 
    \begin{equation}
	\bm{a}_{r}(\phi_{l})=\left[1,e^{j2\pi\frac{d_R}{\lambda}\sin(\phi_{l})},\cdots,e^{j2\pi\frac{(M_r-1)d_R}{\lambda}\sin(\phi_{l})}\right]^T,
   \end{equation}
    where $d_R$ is the distance between any two adjacent antennas.
    
    To simplify analysis, we adopt an observation interval at the ISAC receiver, whose length is larger than the total duration of one time slot and the maximum delay of all the backscattered signals, i.e., $\Delta T + \max\left\{\tau_{0,1},\tau_{0,2},\cdots,\tau_{0,L}\right\}$, to guarantee that the excitation signal $\bm{x}(t)$ is completely included in the backscattered signal from all BDs. By adopting a fixed sampling interval $\Delta t$, which is sufficient small to approximate the transmission delays in the discrete-time domain as integers \cite{Steven}, we discretize the received signals in \eqref{sensing}. Therefore, a set of $M$ vector samples, denoted as $\bm{y}[m]=\bm{y}(m\Delta t),m\in \left\{1,\ldots,M\right\}$, are obtained to represent the received baseband signals at the ISAC receiver, where $M$ is calculated as $M= N +\max\left\{\tau_{0,1},\tau_{0,2},\cdots,\tau_{0,L}\right\}/\Delta t$. Specifically, the $m$-th vector sample obtained from \eqref{sensing} is given as
    \begin{equation}
    \begin{split}
    \label{sample}
    \bm{y}[m]&=\sum_{l=1}^{L}s_l\alpha_l\bm{H}_{l}\bm{x}[m-n_{\tau_{l}}]+\bm{c}[m]+\bm{z}[m], \\
    &=  \bm{H}\bm{x}_{\bm{\tau}}[m]+ \bm{e}[m],
    \end{split}
    \end{equation}
    where $n_{\tau_{l}}=\tau_{0,l}/\Delta t$ denotes the integer transmission delay in the discrete-time domain, $\bm{c}[m]=\bm{c}(m\Delta t)$ and $\bm{z}[m]=\bm{z}(m\Delta t)$ are the vector samples of the clutter and noise, respectively, $\bm{H}=[s_1\alpha_1\bm{H}_{1},s_2\alpha_2\bm{H}_{2},\cdots,s_L\alpha_L\bm{H}_{L}]\in\mathbb{C}^{M_r\times LM_t}$, $\bm{x}_{\bm{\tau}}[m]=\mbox{vec}\left(\bm{X_{\tau}}[m]\right)$, and $\bm{e}[m]=\bm{c}[m]+\bm{z}[m]$. Here, $\bm{X_{\tau}}[m]$ is defined as  $\bm{X_{\tau}}[m]=\big[\bm{x}[m-n_{\tau_{1}}],\bm{x}[m-n_{\tau_{2}}],\cdots,\bm{x}[m-n_{\tau_{L}}]\big]\in\mathbb{C}^{M_t\times L}$, with $\bm{x}[m-n_{\tau_{l}}]=\bm{x}(m\Delta t-\tau_{0,l})$.
    
    \begin{Remark}
    	Asynchrony in ISAC systems may result in various issues, such as measurement ambiguity and accuracy degradation \cite{Problem}, and thus synchronization between the transmitter and the receivers is crucial. In the considered ISAC system, synchronization between the ISAC transmitter and ISAC receivers can be achieved by two methods: (1) Global positioning system (GPS)-based synchronization: the leading node broadcasts a reference clock that is then locked by the other nodes to synchronize with each other\cite{GPS}; and (2) ultra-wide-band (UWB)-based synchronization: data packets with high-resolution timestamps are exchanged among all nodes to achieve synchronization\cite{UWB}. Since we consider the scenario that the ISAC transmitter acts as a high-power node and sends excitation signals towards the designated target area, synchronization can be achieved by the first method.
    \end{Remark}      

\section{Performance Analysis for Sensing and Communication}
\label{Performance}  

In this section, we first introduce a geometric model for sensing (localization), providing a detailed explanation of the motivation for estimating the transmission delay and DoA of the backscattered signals. Then, we derive the FIM with respect to these target parameters based on the received signal described in \eqref{sample}, and accordingly obtain the CRB as the sensing performance metric of the considered system. Finally, we calculate the sum transmission rate of all BDs, thereby providing an assessment of the communication performance of the considered system.

\subsection{Geometric Model for Localization}

Here, we define the unbiased estimators for the transmission delay and DoA given in \eqref{sensing} as $\left\{\widehat{\tau}_{0,l}\right\}_{l=1}^{L}$ and $\left\{\widehat{\phi}_{l}\right\}_{l=1}^{L}$, respectively. Then, by utilizing the unbiased estimators $\left\{\widehat{\tau}_{0,l}\right\}_{l=1}^{L}$ and $\left\{\widehat{\phi}_{l}\right\}_{l=1}^{L}$, along with the geometric model depicted in Fig. \ref{Geometric}, we are able to estimate the location of each BD. Specifically, the distance between the ISAC receiver and the $l$-th BD is calculated by the following proposition.
\begin{figure}[t]
	\centering
	\includegraphics[scale=0.62]{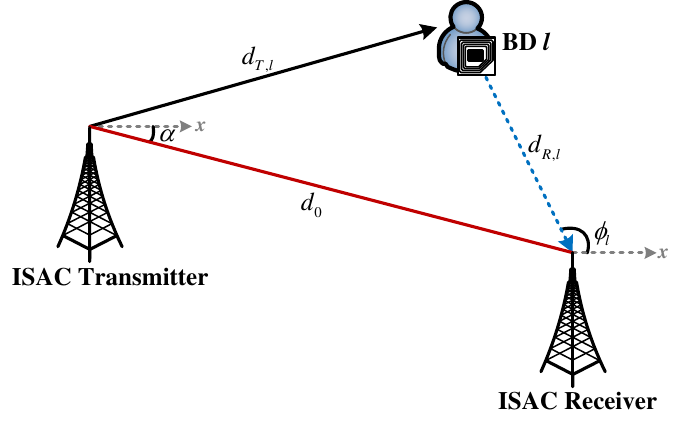}
	\caption{Geometric model for estimating the location of the $l$-th BD via the ISAC transmitter and receiver.}
	\label{Geometric}
\end{figure}
\begin{Proposition}
	\label{location}
	By utilizing the unbiased estimators $\widehat{\tau}_{0,l}$ and $\widehat{\phi}_{l}$, we obtain a closed-form expression of the distance between the ISAC receiver and the $l$-th BD, i.e.,
	 \begin{equation}
	 \label{distance_Pro}
	 \widehat{d}_{R,l}=\frac{c^2(\widehat{\tau}_{0,l}-\tau_{0})^2-d_{0}^2}{2\left[d_{0}\cos(\widehat{\phi}_{l}-\beta)+c(\widehat{\tau}_{0,l}-\tau_{0})\right]},
	 \end{equation}
	 where $\beta$ is the angle between the ISAC transmitter and the ISAC receiver and is calculated as 
	 \begin{equation}
	 \beta =\arctan\left(\frac{y_R-y_T}{x_R-x_T}\right).
	 \end{equation}
\end{Proposition}
\indent \indent \emph{Proof:}
        By applying the Law of Cosine to the triangle formed by the ISAC transmitter, the ISAC receiver, and the $l$-th BD, and utilizing the relationship $d_{T,l}+d_{R,l}=c(\tau_{0,l}-\tau_{0})$ discussed in \eqref{sensing}, the distance between the ISAC receiver and the $l$-th BD is derived. \QEDA

In accordance with the Proposition \ref{location}, the location of the $l$-th BD is then given as
\begin{equation}
\widehat{x}_l = x_R + \widehat{d}_{R,l}\cos(\widehat{\phi}_{l}),
\end{equation}
\begin{equation}
\widehat{y}_l = y_R + \widehat{d}_{R,l}\sin(\widehat{\phi}_{l}).
\end{equation}
Consequently, the DoD at the ISAC transmitter with respect to the $l$-th BD is calculated as 
\begin{equation}
\widehat{\theta_{l}}=\arctan\left(\frac{\widehat{y}_l-y_T}{\widehat{x}_l-x_T}\right).
\end{equation}

\subsection{Localization Performance}
Let $\bm{\rho}\in\mathbb{R}^{2L\times1}$ denote the vector parameter to be estimated at the ISAC receiver, which is defined as
\begin{equation}
\label{target}
\bm{\rho}=[\bm{\tau}^T,\bm{\Phi}^T]^T,
\end{equation}
with $\bm{\tau}=\left[\tau_{0,1},\tau_{0,2},\cdots,\tau_{0,L}\right]^T\in\mathbb{R}^{L\times1}$ and $\bm{\Phi}=[\phi_{1},\phi_{2},\cdots,\phi_{L}]^T\in\mathbb{R}^{L\times1}$. Furthermore, it is evident that the received signal in \eqref{sample} is closely associated with the vector parameter $\bm{\rho}$. Hence, we define a noise-free signal vector at the $m$-th sample in \eqref{sample} as $\bm{\mu}[m,\bm{\rho}]=\bm{H} \bm{x}_{\bm{\tau}}[m]$, and concatenate the received signal $\bm{y}[m]$ in \eqref{sample} into a single column vector, i.e.,  
\begin{equation}
\label{Final}
\bm{y}_s=\bm{\mu}(\bm{\rho}) + \bm{\epsilon},
\end{equation}
where $\bm{y}_s=\mbox{vec}(\bm{Y})\in\mathbb{C}^{M_rM\times1}$ with $\bm{Y}=\big[\bm{y}[1],\bm{y}[2],\cdots,\bm{y}[M]\big]\in\mathbb{C}^{M_r\times M}$, $\bm{\mu}(\bm{\rho})=\mbox{vec}(\bm{U}(\bm{\rho}))$\quad$\in\mathbb{C}^{M_rM\times1}$ with $\bm{U}(\bm{\rho})=\big[\bm{\mu}[1,\bm{\rho}],\bm{\mu}[2,\bm{\rho}],\cdots,\bm{\mu}[M,\bm{\rho}]\big]\in\mathbb{C}^{M_r\times M}$, and   $\bm{\epsilon}=\mbox{vec}(\bm{E})$ with $\bm{E}=\big[\bm{e}[1],\bm{e}[2],\cdots,\bm{e}[M]\big]\sim\mathcal{CN}(0,\bm{C})$. Here, covariance matrix $\bm{C}$ is defined as $\bm{C}=(\sigma^2_{c}+\sigma^2_{z})\bm{I}_{M_r M}$. Hence, FIM with respect to vector parameter $\bm{\rho}$, denoted as $\bm{J}_{\bm{\rho}}\in\mathbb{R}^{2L\times2L}$, is expressed as \cite{FIM}
\begin{equation}
\label{definition}
\bm{J}_{\bm{\rho}}=\mathbb{E}_{\bm{y}_s|\bm{\rho}}\left\{\left(\frac{\partial}{\partial \bm{\rho}}\mbox{ln} \bm{f}(\bm{y}_s|\bm{\rho})\right)\left(\frac{\partial}{\partial \bm{\rho}}\mbox{ln} \bm{f}(\bm{y}_s|\bm{\rho})\right)^T\right\},
\end{equation}
where $\bm{f}(\bm{y}_s|\bm{\rho})$ is the conditional probability density function of observation $\bm{y}_s$. Here, based on the stacked signal vector in \eqref{Final}, $\bm{f}(\bm{y}_s|\bm{\rho})$ takes the following form
\begin{equation}
\begin{split}
	\label{conditional}
\bm{f}(\bm{y}_s|\bm{\rho})&=\frac{1}{\pi^{M_rM}\mbox{det}(\bm{C})}\times\\
&\quad\quad\mbox{exp}\left\{-\left[\bm{y}_s-\bm{\mu}(\bm{\rho})\right]^H\bm{C}^{-1}\left[\bm{y}_s-\bm{\mu}(\bm{\rho})\right]\right\}.
\end{split}
\end{equation}
Based on \eqref{definition} and \eqref{conditional}, we have the following proposition to derive FIM $\bm{J}_{\bm{\rho}}$.
\begin{Proposition}
	\label{Propositon_FIM}
	By taking the first-order partial derivatives of \eqref{conditional}, FIM $\bm{J}_{\bm{\rho}}$ defined in \eqref{definition} can be equivalently rewritten as a block matrix form, i.e., 
	\begin{equation}
		\label{FIM}
		\bm{J}_{\bm{\rho}}=\begin{bmatrix}
			\bm{G}_{\bm{\tau},\bm{\tau}} & \bm{G}_{\bm{\tau},\bm{\Phi}} \\ 
			\bm{G}^T_{\bm{\tau},\bm{\Phi}} & \bm{G}_{\bm{\Phi},\bm{\Phi}} \\ 
		\end{bmatrix},
	\end{equation}
	where each element in \eqref{FIM} is given as
	\begin{equation}
	\label{Element}
			\bm{J}_{\bm{\rho}}[i,j]=\frac{2}{\sigma_{c}^2+\sigma_{z}^2}\, \mathbb{E}\left\{\mathcal{R}\left[\sum_{m=1}^{M}\frac{\partial\bm{\mu}[m,\bm{\rho}]^H}{\partial\rho_{i}}\frac{\partial\bm{\mu}[m,\bm{\rho}]}{\partial\rho_{j}}\right]\right\},
	\end{equation}
	with $\bm{J}_{\bm{\rho}}[i,j]$ being the $(i,j)$-th element of FIM $\bm{J}_{\bm{\rho}}$ and $\rho_{i}$ being the $i$-th element of the vector parameter $\bm{\rho}$, i.e., $\tau_{0,l}$ and $\phi_{l}$, $\forall i,j\in[2L]$. By substituting the first-order partial derivatives $\frac{\partial\bm{\mu}[m,\bm{\rho}]}{\partial\tau_{0,l}}$ and $\frac{\partial\bm{\mu}[m,\bm{\rho}]}{\partial\phi_{l}}$ into \eqref{Element}, the closed-form expressions of components $\bm{G}_{\bm{\tau},\bm{\tau}}$, $\bm{G}_{\bm{\tau},\bm{\Phi}}$, and $\bm{G}_{\bm{\Phi},\bm{\Phi}}$ are respectively expressed as 
		\begin{equation}
		\label{partone}
		\bm{G}_{\bm{\tau},\bm{\tau}}= \varepsilon_{g}\overline{F_g^2}\,\bm{\xi} \cdot \bm{B}_{\bm{\tau},\bm{\tau}} , 
	\end{equation}
	\begin{equation}
		\label{parttwo}
		\bm{G}_{\bm{\tau},\bm{\Phi}}= \bm{\xi} \cdot \bm{B}_{\bm{\tau},\bm{\Phi}}, 
	\end{equation}
	and 
	\begin{equation}
		\label{partthree}
		\bm{G}_{\bm{\Phi},\bm{\Phi}}= \varepsilon_{g}\,\bm{\xi} \cdot \bm{B}_{\bm{\Phi},\bm{\Phi}},
	\end{equation}
	where $\overline{F_g^2}$ is the mean square bandwidth of the transmit pulse function $g(t)$ and is defined as $\overline{F_g^2}=\frac{\int_{0}^{\Delta t}|\dot{g}(t)|^2 dt}{\int_{0}^{\Delta t} |g(t)|^2 dt}$, with $\dot{g}(t)=\frac{\partial g(t)}{\partial t}$ being the first-order derivative of $g(t)$ with respect to $t$; $\bm{\xi}$, $\bm{B}_{\bm{\tau},\bm{\tau}}$, $\bm{B}_{\bm{\tau},\bm{\Phi}}$, and $\bm{B}_{\bm{\Phi},\bm{\Phi}}$ are all diagonal matrices, and their $l$-th diagonal elements are expressed as $\xi_l=\frac{2\varrho Ns_{l}^2 \eta(d_{T,l})\eta(d_{R,l})}{(\sigma_{c}^2+\sigma_{z}^2)\Delta t}$, $\bm{B}_{\bm{\tau},\bm{\tau}}[l,l]=\mbox{Tr}(\bm{H}_l\bm{R}_x\bm{H}_l^H)$,
	$\bm{B}_{\bm{\tau},\bm{\Phi}}[l,l]=\mathcal{R}\left[\varepsilon_{\dot{g}}\mbox{Tr}(\bm{H}_l\bm{R}_x\bm{H}_l^H\bm{\Lambda}^H)\right]$, and $\bm{B}_{\bm{\Phi},\bm{\Phi}}[l,l]=\mbox{Tr}(\bm{\Lambda}\bm{H}_l\bm{R}_x\bm{H}_l^H\bm{\Lambda}^H)$, respectively, with $\bm{\Lambda}=\mbox{diag}\left\{0,1,\cdots,M_r-1\right\}$ and $\varepsilon_{\dot{g}}=\int_{0}^{\Delta t} (\dot{g}(t))^{*}g(t) dt$.
\end{Proposition}
\indent \indent \emph{Proof:}
Please see Appendix \ref{AppendixA}.\QEDA

By using Proposition \ref{Propositon_FIM}, we obtain FIM with respect to vector parameter $\bm{\rho}$. To theoretically analyze the localization performance, we adopt CRB as the performance metric of the considered system. Specifically, the CRB matrix for estimating vector parameter $\bm{\rho}$ is defined as the inverse of FIM $\bm{J}_{\bm{\rho}}$, i.e., $\bm{J}_{\bm{\rho}}^{-1}$, and the CRB on vector parameter $\bm{\rho}$, denoted as CRB$(\bm{\rho})$, is defined as the trace of the CRB matrix, i.e.,
\begin{equation}
	\label{CRBmatrix}
	\mbox{CRB}(\bm{\rho})=\mbox{Tr}\left(\bm{J}_{\bm{\rho}}^{-1}\right).
\end{equation}
We then have the following proposition to derive the closed-form expression for the CRB on vector parameter $\bm{\rho}$.
\begin{Proposition}
\label{CRBMatrix}
The closed-form expression for CRB on vector parameter $\bm{\rho}$ is given as 
\begin{equation}
\label{CRB}
	\text{CRB}(\bm{\rho})=  \text{CRB}(\bm{\tau})+\text{CRB}(\bm{\Phi}),
\end{equation}
where $\mbox{CRB}(\bm{\tau})$ and $\mbox{CRB}(\bm{\Phi})$ are the closed-form expressions for the CRB on the transmission delay $\bm{\tau}$ and DoA $\bm{\Phi}$, respectively, and are calculated as  
\begin{equation}
\label{CRB-1}
\text{CRB}(\bm{\tau})=\sum_{l=1}^{L} \frac{\varepsilon_{g}h_l(\bm{R}_x)}{\varepsilon_{g}^2\overline{F_g^2}f_l(\bm{R}_x)h_l(\bm{R}_x)-|\varepsilon_{\dot{g}}|^2g_l^2(\bm{R}_x)}\times\frac{1}{\xi_l},
\end{equation}
and
\begin{equation}
\label{CRB-2}
	\mbox{CRB}(\bm{\Phi})=\sum_{l=1}^{L}\frac{\varepsilon_{g}\overline{F_g^2}f_l(\bm{R}_x)}{\varepsilon_{g}^2\overline{F_g^2}h_l(\bm{R}_x)f_l(\bm{R}_x)-|\varepsilon_{\dot{g}}|^2g_l^2(\bm{R}_x)}\times\frac{c_l}{\xi_l}, 
\end{equation}
respectively, with $c_l=\frac{\lambda^2}{4\pi^2 d_R^2\cos(\phi_{l})^2}$, $f_l(\bm{R}_x)=\mbox{Tr}(\bm{H}_l\bm{R}_x\bm{H}_l ^H)$, $g_l(\bm{R}_x)=\mbox{Tr}(\bm{H}_l\bm{R}_x\bm{H}_l^H\bm{\Lambda}^H)$, and $h_l(\bm{R}_x)=\mbox{Tr}(\bm{\Lambda}\bm{H}_l\bm{R}_x\bm{H}_l^H\bm{\Lambda}^H)$.
\end{Proposition}
\indent \indent \emph{Proof:}
	Please see Appendix \ref{AppendixD}.\QEDA

Observing from \eqref{CRB-1} and \eqref{CRB-2}, it is easy to obtain that the CRB on the transmission delay and DoA is a function of the sample covariance matrix $\bm{R}_x$.

\subsection{Communication Performance}

By considering the fact that transmission delay of the backscattered signals is significantly smaller than the coherence time of each transmission block\footnote{It always holds since BackCom systems are typically deployed in short-range scenarios\cite{MASK}.}, we ignore the transmission delays in \eqref{sample} to simplify the analysis in this subsection. Consequently, this procedure yields a vector signal from \eqref{sample}, denoted as $\widetilde{\bm{y}}[n]$$\in\mathbb{C}^{M_r\times1}$, which is expressed as
\begin{equation}
\label{result}
	\bm{\widetilde{y}}[n]=\sum_{l=1}^{L}\alpha_l\bm{H}_l\bm{x}[n]s_l +\bm{e}[n].
\end{equation}
Next, to facilitate the following analysis, we concatenate signal $\bm{\widetilde{y}}[n]$ in \eqref{result} into a single column vector, i.e.,
\begin{equation}
\label{multiple}
	\bm{y}_c=\sum_{l=1}^{L}\bm{w}_ls_l+\bm{\epsilon},
\end{equation}
where $\bm{y}_c=\mbox{vec}(\bm{\widetilde{Y}})\in\mathbb{C}^{M_rN\times1}$ with $\bm{\widetilde{Y}}=\big[\bm{\widetilde{y}}[1],\bm{\widetilde{y}}[2],\cdots,\bm{\widetilde{y}}[N]\big]\in\mathbb{C}^{M_r\times N}$, and $\bm{w}_l=\mbox{vec}(\bm{W}_l)\in\mathbb{C}^{M_rN\times1}$ is the channel gain of the $l$-th BD, with $\bm{W}_l=\alpha_l\bm{H}_l\bm{X}$ and $\bm{X}=\big[\bm{x}[1],\bm{x}[2],\cdots,\bm{x}[N]\big]\in\mathbb{C}^{M_t\times N}$. 

The input-output relationship between the BDs and the ISAC receiver, as described in \eqref{multiple}, is modeled as a multiple access channel with $L$ users \cite{TDM,AddTwo}, and thus the sum transmission rate of all BDs is adopted as the communication performance metric of the considered system. Under the multiple access scenario, the sum transmission rate of all BDs is given as \cite{neglect2}
\begin{equation}
	\label{capacity}
	C_{\text{sum}}=\frac{1}{N}\mbox{log}\,\mbox{det}\left(\bm{I}_{M_rN}+\frac{1}{\sigma_{c}^2+\sigma_{z}^2}\sum_{l=1}^{L}(s_l\bm{w}_l)(s_l\bm{w}_l)^H\right).
\end{equation}
where the pre-log factor $1/N$ accounts for the fact that the symbol period of each BD is $N$ times greater than that of the excitation signal.

By introducing an auxiliary matrix,  $\widetilde{\bm{W}}=\left[s_1\bm{w}_1,s_2\bm{w}_2,\cdots,s_L\bm{w}_L\right]$, we rewrite the sum transmission rate in \eqref{capacity} of all BDs as an equivalent form, i.e.,
\begin{align}
	\label{rate}
		C
		_{\text{sum}}&=\frac{1}{N} \text{log}\,\text{det}\left(\bm{I}_{M_rN}+\frac{1}{\sigma_{c}^2+\sigma_{z}^2}\widetilde{\bm{W}}\widetilde{\bm{W}}^H\right), \notag \\
		&\overset{(a)}{=}\frac{1}{N} \text{log}\,\text{det}\left(\bm{I}_{L}+\frac{1}{\sigma_{c}^2+\sigma_{z}^2}\widetilde{\bm{W}}^H\widetilde{\bm{W}}\right), \notag \\
		&\overset{(b)}{=}\frac{1}{N} \text{log}\;\text{det} \left(\bm{I}_L+\frac{N}{\sigma_{c}^2+\sigma_{z}^2}\bm{F}\right),
	\end{align}
	where  the $(i,j)$-th element of $\bm{F}$ is denoted as $\bm{F}[i,j]=\text{Tr}\left(s_i^Hs_j\alpha_i^H\alpha_j\bm{H}_j\bm{R}_x\bm{H}_i^{H}\right)$; equality (a) is obtained by Weinstein-Aronszajn identity $\text{det}\left(\bm{I}_m+\bm{A\bm{B}}\right)=\text{det}\left(\bm{I}_n+\bm{B\bm{A}}\right)$\cite{Weinstein}; equality (b) is obtained by  $\left(\mbox{vec}(\bm{A})\right)^H\mbox{vec}(\bm{A})=\mbox{Tr}(\bm{A}^H\bm{A})$, $\mbox{Tr}(\bm{A}\bm{B})=\mbox{Tr}(\bm{B}\bm{A})$, and $\bm{X}\bm{X}^H=\sum_{n=1}^{N}\bm{x}[n]\bm{x}[n]^H=N\bm{R}_x$.

\subsection{Discussion}

The CRB for parameter estimations in \eqref{CRB} and the sum transmission rate of all BDs in \eqref{rate} are both functions of the sample covariance matrix $\bm{R}_x$, as the excitation signal $x(t)$ is reused for performing localization and communication tasks in our considered setup. Furthermore, an inherent conflict exists between these two tasks \cite{ISAC2},\cite{clutter}, since they share wireless resources within one integrated hardware system, leading to the potential mutual interference between localization and communication signals \cite{CRB-single}. Therefore, there exists a trade-off between the localization and communication tasks.

\section{Transmit Beamforming for CRB Minimization}
\label{Optimization}

We are particularly interested in revealing the trade-off between the CRB for localization and the sum transmission rate for communication. To this end, we define the achievable performance region $\mathcal{C}(\Psi,\Gamma)$ of the considered ISAC system, subject to the constraint of power budget, as
\begin{align}
	\mathcal{C}(\Psi,\Gamma)=\big\{(\Psi,\Gamma):&\Psi\geq\mbox{CRB}(\bm{\rho}),\notag\\&\,\Gamma\leq C_{\text{sum}},\, p_g\mbox{Tr}(\bm{R}_x)\leq P_0\big\},
\end{align}
where $(\Psi,\Gamma)$ is the CRB-rate pair that can be simultaneously achieved by the considered system under the given power budget.

\vspace{-2pt}
To characterize the boundary of the achievable performance region, we propose a method to minimize the CRB on the transmission delay and DoA by designing the sample covariance matrix $\bm{R}_x$ at the ISAC transmitter, while subject to the transmission power constraint at the ISAC transmitter and the sum transmission rate constraint for all BDs. Then, the CRB minimization problem is formulated as 
\begin{eqnarray}
\label{objective}
& \min\limits_{ \{\bm{R}_x \} }   & \text{CRB}(\bm{\rho})  \\
\label{constraint-one}
& \text{s.t.} &  p_{g}\text{Tr}(\bm{R}_x)\leq P_0,  \\
\label{constraint-two}
&\quad & C_{\text{sum}}\geq \Gamma_{\text{th}},
\end{eqnarray}
where objective function \eqref{objective} is the CRB derived in \eqref{CRB}; \eqref{constraint-one} is the transmission power constraint at the ISAC transmitter, with $P_0$ being the power budget; and \eqref{constraint-two} is the sum transmission rate constraint of all BDs, with $\Gamma_{\text{th}}$ being the minimum sum transmission rate. 

\vspace{-3pt}
In Proposition \ref{CRBMatrix}, it is demonstrated that the CRB on the transmission delay and DoA of the backscattered signals from the $l$-th BD, i.e., $\text{CRB}(\tau_{l})$ and $\text{CRB}(\phi_{l})$, are composite functions of the design variable $\bm{R}_x$. It is easy to obtain see problem \eqref{objective}-\eqref{constraint-two} is non-convex due to the coupling of inner functions $f_l(\bm{R}_x)$, $g_l(\bm{R}_x)$, and $h_l(\bm{R}_x)$ within $\text{CRB}(\tau_{l})$ and $\text{CRB}(\phi_{l})$. Fortunately, since CRB serves as the lower limit on the variance of any unbiased estimator \cite{Steven}, it is easy to check that $\text{CRB}(\tau_l)\geq0$ and $\text{CRB}(\phi_l)\geq0$ are always valid. Additionally, functions $f_l(\bm{R}_x)$ and $h_l(\bm{R}_x)$ remain positive for all possible values of $\bm{R}_x$. These two inherent properties determine that both the numerator and the denominator of $\text{CRB}(\tau_{l})$ and $\text{CRB}(\phi_{l})$ are strictly positive. Hence, problem \eqref{objective}-\eqref{constraint-two}  manifests as a sum-of-functions-of-ratios problem \cite{FP}, which can be effectively addressed by utilizing FP techniques \cite{FP}. To effectively solve problem \eqref{objective}-\eqref{constraint-two}, we propose an approach that combines FP \cite{FP} and Schur complement techniques \cite{Schur} to transform it into an equivalent convex form. The detailed process of the proposed approach is discussed as follows.

\vspace{-3pt}
First, by applying Theorem 2 in \cite{FP}, we reformulate problem \eqref{objective}-\eqref{constraint-two} into a standard form of FP problem as presented in \cite{FP}, i.e.,
\begin{eqnarray}
\label{objective-FP}
&\max\limits_{ \{\bm{R}_x \} }   & \sum_{l=1}^{L}P_{l}^{+}
\bigg(\frac{1}{\text{CRB}(\tau_l)}\bigg) + \sum_{l=1}^{L}Q_{l}^{+}\bigg(\frac{1}{\text{CRB}(\phi_l)}\bigg)\\
\label{constraint-two-one}
& \text{s.t.} & \eqref{constraint-one},\eqref{constraint-two}     ,
\end{eqnarray}
where $P_{l}^{+}(x)=-1/x$ and $Q_{l}^{+}(x)=-1/x$, $\forall l\in \mathcal{L}$, are concave increasing functions. However, the non-concavity of inner functions $\frac{1}{\text{CRB}(\tau_{l})}$ and $\frac{1}{\text{CRB}(\phi_{l})}$ within functions $P_l^{+}(\cdot)$ and $Q_l^{+}(\cdot)$ leads to the non-convexity of objective function $\eqref{objective-FP}$. Then, to effectively address this challenge, we introduce two distinct sets of auxiliary variables $\bm{\omega}=\left\{\omega_l,\omega_2,\cdots,\omega_L\right\}$ and $\bm{\nu}=\left\{\nu_l,\nu_2,\cdots,\nu_L\right\}$ to transform problem \eqref{objective-FP}-\eqref{constraint-two-one} into an equivalent form, i.e., 
\begin{align}
	\label{objective-mid}
	\max\limits_{ \{\bm{R}_x, \bm{\omega},\bm{\nu} \} }   & \sum_{l=1}^{L}P_{l}^{+}
	(\omega_l) + \sum_{l=1}^{L}Q_{l}^{+}(\nu_l)\\
	\label{constraint_Schur_one}
	 \text{s.t.}\,\quad & \varepsilon_{g}\overline{F_g^2}f_l(\bm{R}_x)-\frac{|\varepsilon_{\dot{g}}|^2g_l^2(\bm{R}_x)}{\varepsilon_{g}h_l(\bm{R}_x)} - \frac{\omega_l}{\xi_l} \geq 0, \;\forall l\in\mathcal{L},                 \\
	\label{constraint_Schur_two}
	 \quad &  \varepsilon_{g}h_l(\bm{R}_x)-\frac{|\varepsilon_{\dot{g}}|^2g_l^2(\bm{R}_x)}{\varepsilon_{g}\overline{F_g^2}f_l(\bm{R}_x)}-\frac{c_l\nu_l}{\xi_l}\geq 0,  \;\forall l\in\mathcal{L},\\
	 \label{constraint_Schur_three}
	 \quad & \eqref{constraint-one},\eqref{constraint-two},
\end{align}
where $\xi_l$, $f_l(\bm{R}_x)$, $g_l(\bm{R}_x)$, and $h_l(\bm{R}_x)$ are defined in Proposition \ref{CRBMatrix}. Finally, by applying Schur complement technique \cite{Schur}, we transform constraints in \eqref{constraint_Schur_one} and \eqref{constraint_Schur_two} into the following convex semi-definite constraints, i.e.,
\begin{equation}
\label{final-one}
\begin{bmatrix}
	\varepsilon_{g}\overline{F_g^2}f_l(\bm{R}_x)-\frac{\omega_l}{\xi_l} & |\varepsilon_{\dot{g}}|g_l(\bm{R}_x)\\
	|\varepsilon_{\dot{g}}|g_l(\bm{R}_x) & \varepsilon_{g}h_l(\bm{R}_x)
\end{bmatrix}\succeq 0,\;\forall l\in\mathcal{L}, 
\end{equation}
\begin{equation}
\label{final-two}
\begin{bmatrix}
	\varepsilon_{g}h_l(\bm{R}_x)-\frac{c_l\nu_l}{\xi_l} & |\varepsilon_{\dot{g}}|g_l(\bm{R}_x)\\
	|\varepsilon_{\dot{g}}|g_l(\bm{R}_x) & \varepsilon_{g}\overline{F_g^2}f_l(\bm{R}_x)
\end{bmatrix}\succeq 0,\;\forall l\in\mathcal{L}.
\end{equation}

Based on the above analysis, problem \eqref{objective-mid}-\eqref{constraint_Schur_three} is transformed into an equivalent form, i.e.,
\begin{eqnarray}
	\label{objective-Schur}
	& \max\limits_{ \{\bm{R}_x, \bm{\omega},\bm{\nu} \} }   & \sum_{l=1}^{L}P_l^{+}\left(\omega_l\right)+\sum_{l=1}^{L}Q_l^{+}\left(\nu_l\right)   \\
	\label{constraint-Schur-one}
	& \text{s.t.} &   P_0 - p_{g}\mbox{Tr}(\bm{R}_x)\geq 0, \\
	\label{constraint-Schur-two}
	&\quad & C_{\text{sum}} -\Gamma_{\text{th}} \geq 0, \\
	\label{constraint-Schur-three}
	&\quad & \eqref{final-one},\eqref{final-two},
\end{eqnarray}
where \eqref{constraint-Schur-one} and \eqref{constraint-Schur-two} are obtained by rearranging the terms in \eqref{constraint-one} and \eqref{constraint-two}, respectively. It is easy to obtain that objective function \eqref{objective-Schur} and constraints \eqref{constraint-Schur-one}-\eqref{constraint-Schur-three} are convex. Hence, problem \eqref{objective-Schur}-\eqref{constraint-Schur-three} is a convex problem, which can be efficiently solved by some optimization tools, e.g., CVX \cite{CVX}.

\section{Numerical Results}
\label{Simulation}

In this section, we provide numerical results to validate the performance of our proposed transmit beamforming method under typical scenarios, which are denoted as Scenarios I and II in Fig. \ref{Two Scenarios}, i.e, with one single BD and multiple ones distributed within the target area, respectively. As depicted in Fig. \ref{Two Scenarios}, the locations of the ISAC transmitter and the ISAC receiver are set as $(x_T,y_T)=(0 \;\mbox{m} ,0 \;\mbox{m})$ and $(x_R,y_R)=(2\;\mbox{m}, -1.5\;\mbox{m})$, respectively.

For Scenario I, the location of the single BD is set as $(x_1,y_1)=(1.5 \;\mbox{m} ,-0.5 \;\mbox{m})$, and the transmitted information symbol of the single BD within the considered time slot is randomly generated from the modulation alphabet set $\mathcal{S}_{D}$\textbackslash$\{0\}$. For Scenario II, we consider a BackCom system for ISAC with $L=9$ BDs, where all BDs are uniformly distributed in a circle with center $(1.5 \;\mbox{m},-0.5 \;\mbox{m})$ and radius $0.5 \;\mbox{m}$, and the transmitted information symbols of all BDs within the considered time slot of all BDs are randomly generated from the modulation alphabet set $\mathcal{S}_{D}$. Other parameters adopted in these two scenarios are detailed as follows: For the ISAC transmitter and ISAC receiver, the distance between any two adjacent antennas is set as half of the carrier wavelength, i.e., $d_T=d_R=\frac{\lambda}{2}$; for the excitation signal, the symbol duration is set as $\Delta t=5\cdot10^{-7}\;\mbox{s}$, the number of the excitation symbol is set as $N=128$, and the transmit pulse functions are set as $g_1(t)=\sqrt{2}\cos(\frac{\pi}{2\Delta t}t)$, $g_2(t)=\sqrt{\frac{\pi}{a}}\mbox{sinc}(\frac{t}{\Delta t})$, and $g_3(t)=\sqrt{3}(1-\frac{t}{\Delta t})$, respectively, where $a=\int_{0}^{\pi}\frac{\sin^2(x)}{x^2}dx$; for each BD, the fraction factor is set as $\varrho=1$, and the modulation alphabet set $\mathcal{S}_{D}$ is set as $\left\{0,\frac{1}{8},\frac{1}{4},\frac{3}{8},\frac{1}{2},\frac{5}{8},\frac{3}{4},\frac{7}{8},1\right\}$; the path loss at the reference distance of 1 m is set as $\zeta=30\; \mbox{dB}$, and the path loss exponent is set as $\gamma=2.7$ \cite{TransmitBeamforming}; the noise power values are set as $\sigma_{c}^2=\sigma_{z}^2= -60\; \mbox{dBm}$.
 
\begin{figure}[t]
	\centering
	\subfigure[Scenario I: Single BD.]{\includegraphics[scale=0.43]{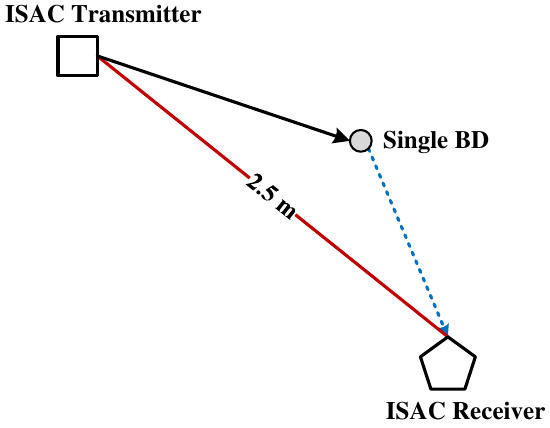}}
	\subfigure[Scenario II: Multiple BDs.]{\includegraphics[scale=0.43]{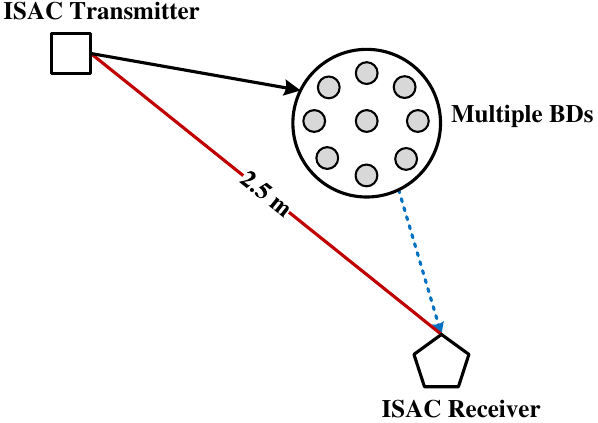}}
	\caption{Experimental scenarios of the considered BackCom system for ISAC.}
	\label{Two Scenarios}      
\end{figure}
 
\begin{figure}[t]
	\centering
	\subfigure[Scenario I.]{\includegraphics[scale=0.50]{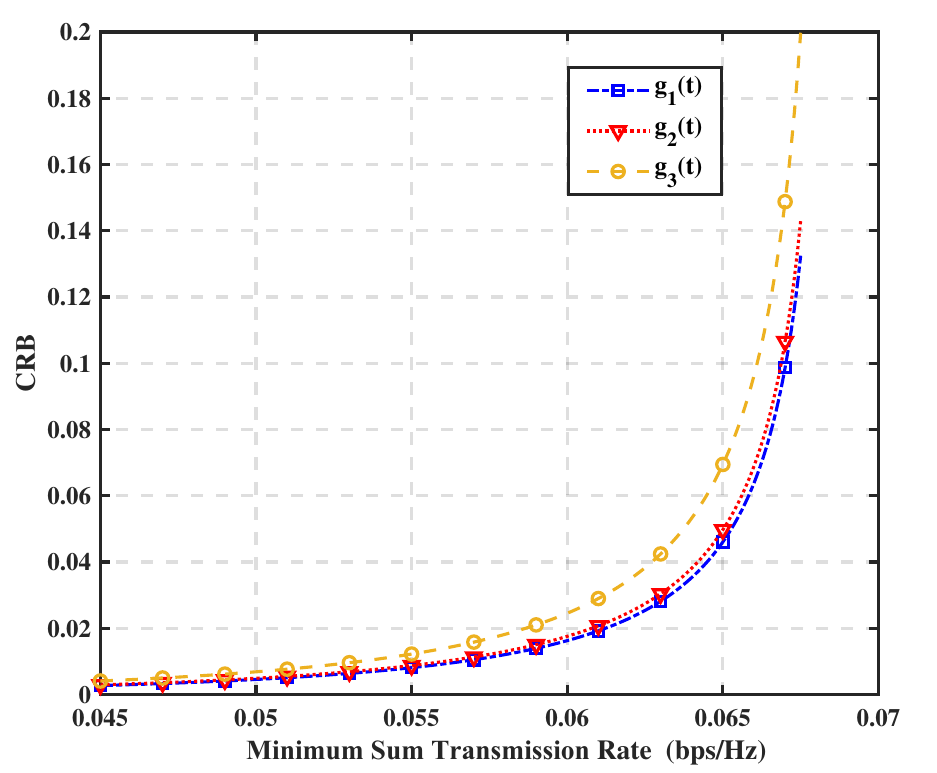}}
	\subfigure[Scenario II.]{\includegraphics[scale=0.50]{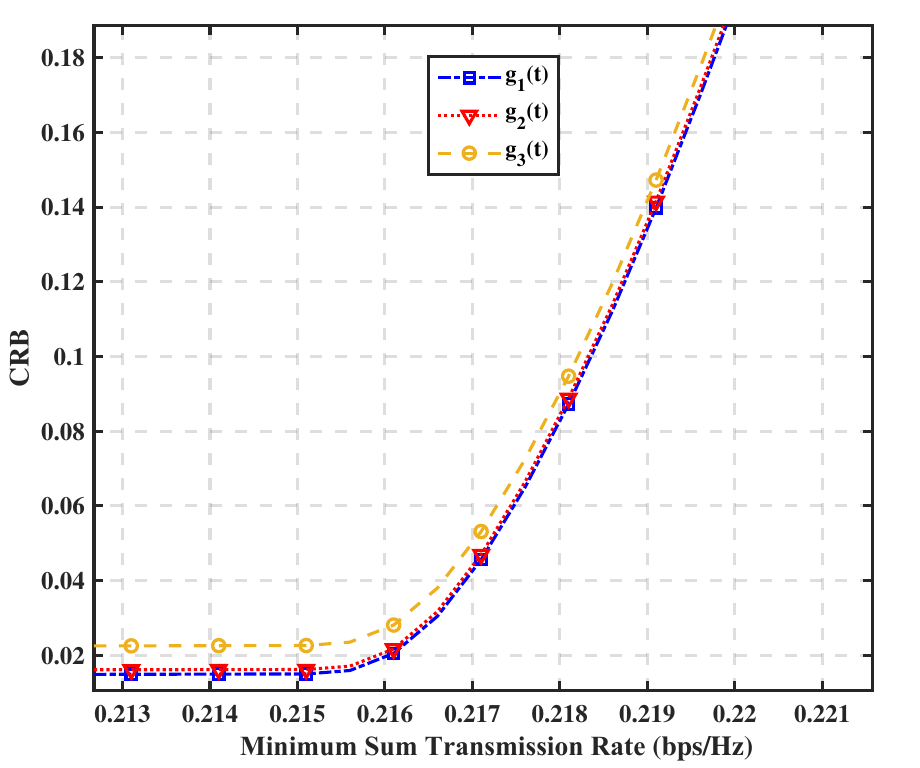}}
	\caption{Trade-off curve between CRB and minimum sum transmission rate with different transmit pulses, with $P_0=30$ dBm and $M_t=M_r=8$.}
	\label{Transmit Pulses}      
\end{figure}

First, we plot the trade-off curve between CRB and minimum sum transmission rate for the proposed method with different transmit pulse functions in the two scenarios, as depicted in Fig. \ref{Transmit Pulses}(a) and Fig. \ref{Transmit Pulses}(b), respectively. The power budget is set as $P_0 = 30\; \mbox{dBm}$, and the number of antennas at the ISAC transmitter and ISAC receiver are set as $M_t=M_r=8$. It is observed that the cosine transmit pulse function $g_1(t)$ consistently outperforms the other two functions over all minimum sum transmission rates. For instance, when the minimum sum transmission rate is fixed at $0.05$ bps/Hz in Scenario I, the CRB with transmit pulse functions $g_1(t)$, $g_2(t)$, and $g_3(t)$ are $4.5\times10^{-3}$, $4.8\times10^{-3}$, and $6.9\times10^{-3}$, respectively. Moreover, these three transmit pulse functions share the same maximum sum transmission rate $0.067$ bps/Hz in Scenario I and $0.227$ bps/Hz in Scenario II. Therefore, in the subsequent simulations, we adopt $g_1(t)$ as the transmit pulse function unless otherwise stated.

\begin{figure}[t]
	\centering
	\includegraphics[scale=0.5]{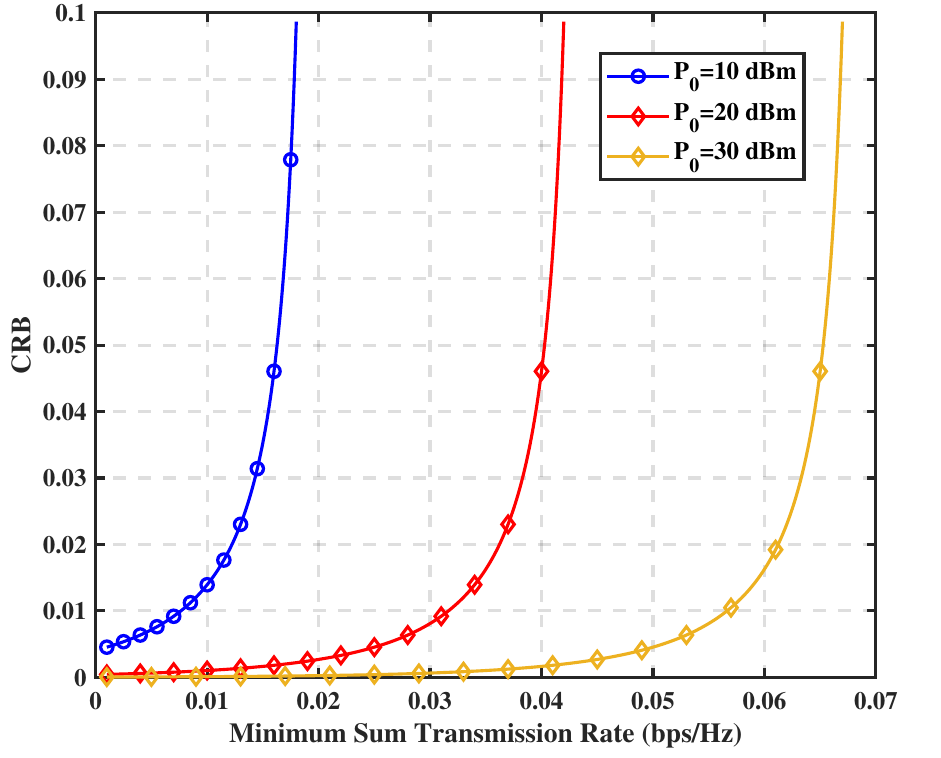}
	\caption{Trade-off curve between CRB and minimum sum transmission rate in Scenario I with different power budgets, with $M_t=M_r=8$.}
	\label{PowerBudget}
\end{figure}

Next, Fig. \ref{PowerBudget} plots the trade-off curve between CRB and minimum sum transmission rate for the proposed method with different power budgets in Scenario I. The number of antennas at the ISAC transmitter and ISAC receiver are set as $M_t=M_r=8$. It is observed that the localization and communication performance both exhibit significant improvements as the power budget increases, since more power is allocated to perform these tasks. For instance, when the minimum sum transmission rate is fixed at $0.005$ bps/Hz, the CRB with power budgets: $10$ dBm, $20$ dBm and $30$ dBm are $7.2\times10^{-3}$, $6.5\times10^{-4}$, and $6.9\times10^{-5}$, respectively, and when the CRB is fixed at $0.02$, the minimum sum transmission rates with power budgets: $10$ dBm, $20$ dBm and $30$ dBm are $0.012$ bps/Hz, $0.036$ bps/Hz, and $0.061$ bps/Hz, respectively. Moreover, it is easy to determine the boundary point that segregates the trade-off curve into dual-constrained and single-constrained segments. For instance, when the power budget is $P_0 = 30\; \mbox{dBm}$ and the minimum sum transmission rate is below $0.005$ bps/Hz, the CRB remains unaffected by the minimum sum transmission rate, remaining at about $6.5\times10^{-4}$. However, when the power budget is $P_0 = 30\; \mbox{dBm}$ and the minimum sum transmission rate exceeds $0.044$ bps/Hz, the problem becomes infeasible, resulting in a failure for the localization task.

Fig. \ref{Antennas} plots the trade-off curve between CRB and minimum sum transmission rate for the proposed method with different number of transmitting and receiving antennas. The power budget is set as $P_0 = 30\; \mbox{dBm}$. It is observed that the number of receiving antennas has a more significant influence on both the localization and communication performance compared to the number of transmitting antennas. This disparity is due to the utilization of spatial diversity at the receiving antennas to mitigate channel fading and noise. For instance, when the minimum sum transmission rate is fixed at $0.04$ bps/Hz, increasing the number of receiving antennas from $M_r=4$ to $M_r=8$ results in a decrease in the CRB from $3.47\times10^{-2}$ to $2.99\times10^{-3}$, while increasing the number of transmitting antennas from $M_t=4$ to $M_t=8$ only leads to a decrease in the CRB from $3.47\times10^{-2}$ to $1.60\times10^{-3}$. Furthermore, when the CRB is fixed at $0.2$, increasing the number of receiving antennas from $M_r=4$ to $M_r=8$ results in a maximum sum transmission rate increase of $0.011$ bps/Hz, while increasing the number of transmitting antennas from $M_t=4$ to $M_t=8$ only yield an increase of $0.007$ bps/Hz.

\begin{figure}[t]
	\centering
	\includegraphics[scale=0.5]{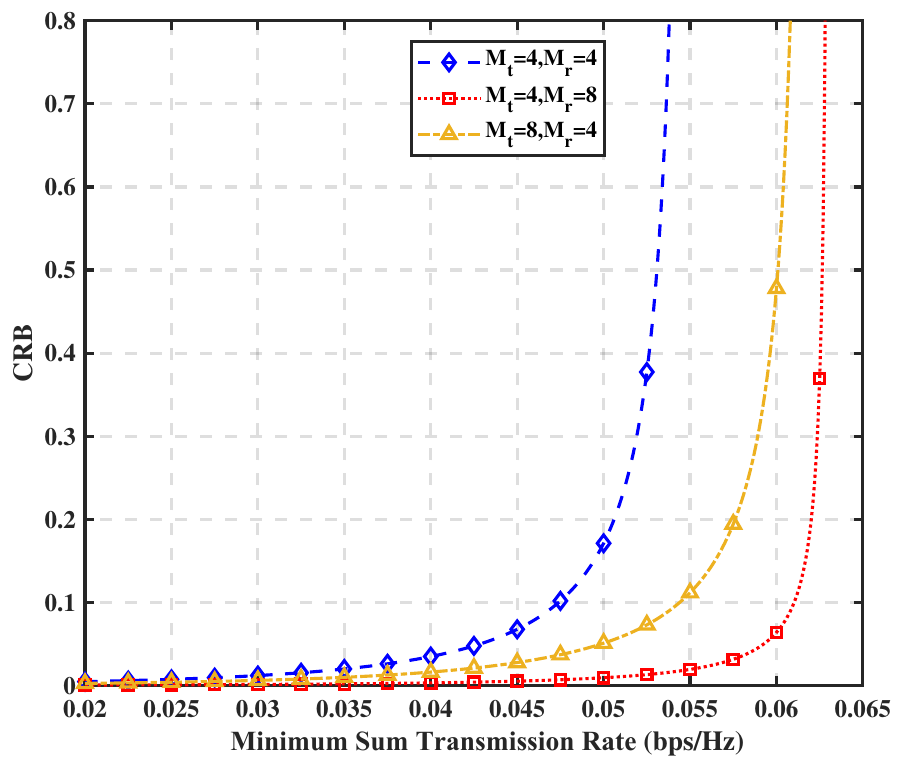}
	\caption{Trade-off curve between CRB and minimum sum transmission rate in Scenario I with different number of antennas, with $P_0=30$ dBm.}
	\label{Antennas}
\end{figure}

\begin{figure}[t]
	\centering
	\includegraphics[scale=0.5]{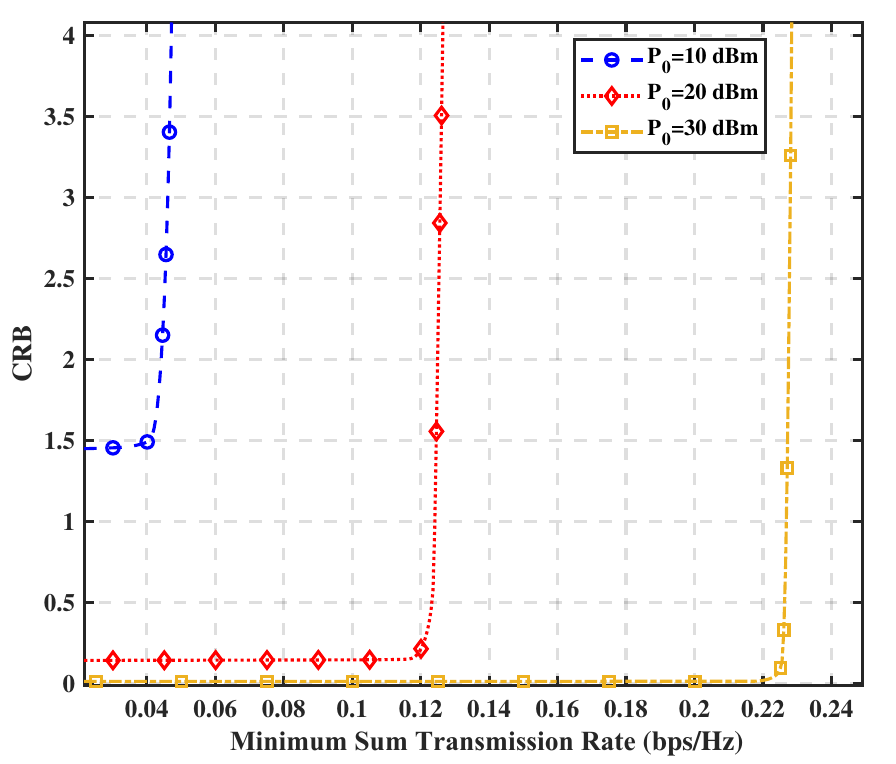}
	\caption{Trade-off curve between CRB and minimum sum transmission rate in Scenario II with different power budgets, with $M_t=M_r=8$.}
	\label{PowerBudget2}
\end{figure}

\begin{figure}[t]
	\centering
	\includegraphics[scale=0.5]{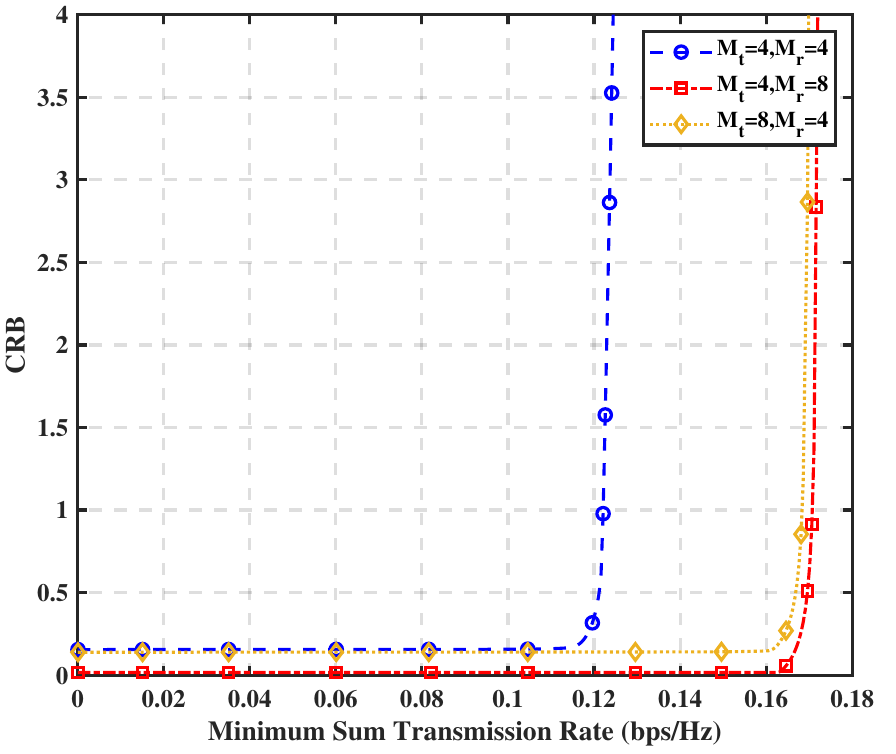}
	\caption{Trade-off curve between CRB and minimum sum transmission rate in Scenario II with different number of antennas, with $P_0=30$ dBm.}
	\label{Antennas2}
\end{figure}

Finally, we plot the trade-off curve between CRB and minimum sum transmission rate for the proposed method with different power budgets and different number of transmitting and receiving antennas in Scenario II, as depicted in Fig. \ref{PowerBudget2} and Fig. \ref{Antennas2}, respectively. The number of antennas at the ISAC transmitter and ISAC receiver are set as $M_t=M_r=8$ in Fig. \ref{PowerBudget2}, and the power budget is set as $P_0=30$ dBm in Fig. \ref{Antennas2}. It is observed that the trade-off curve between CRB and minimum sum transmission rate shown in Fig. \ref{PowerBudget2} and Fig. \ref{Antennas2} are similar to that shown in Fig. \ref{PowerBudget} and Fig. \ref{Antennas}, respectively, and can be similarly explained.

\section{Conclusion}
\label{conclusion}

In this paper, we considered a multi-user BackCom system for ISAC. This system consists of one ISAC transmitter with multiple antennas, multiple single-antenna passive BDs, and one ISAC receiver with multiple antennas, and is capable of simultaneously performing sensing (localization) and communication tasks. Under this scenario, we first derived the closed-form expression on the CRB for parameter estimations and the sum transmission rate of all BDs. Then, we formulated an optimization problem aiming to minimize the obtained CRB by optimizing the sample covariance matrix at the ISAC transmitter, subject to the constraints of sum transmission rate and power budget. Although the formulated problem is non-convex, we proposed an approach that combines FP and Schur complement techniques to equivalently transform it into a convex form. Finally, numerical results presented the trade-off curve between the CRB and minimum sum transmission rate with different parameters.

\begin{appendices}

\section{Proof of Proposition \ref{Propositon_FIM}}
\label{AppendixA}

First, the step-by-step derivation of the \eqref{Element} is presented below. Based on the conditional probability density function in \eqref{conditional}, the log-likelihood function of $\bm{f}(\bm{y}_s|\bm{\rho})$ is expressed as 
\begin{equation}
\begin{split}
\label{log}
\mbox{ln}\bm{f}(\bm{y}_s|\bm{\rho}) =&-M_rM\,\mbox{ln}\pi- \mbox{ln}\;\mbox{det}(\bm{C})\\&- \left[\bm{y}_s-\bm{\mu}(\bm{\rho})\right]^H\bm{C}^{-1}\left[\bm{y}_s-\bm{\mu}(\bm{\rho})\right].
\end{split}
\end{equation} 
The first-order partial derivative of $\mbox{ln}\bm{f}(\bm{y}_s|\bm{\rho})$ with respect to $\rho_{i}$ is calculated as  
\begin{align}
\frac{\partial\,\mbox{ln}\bm{f}(\bm{y}_s|\bm{\rho})}{\partial\rho_{i}}
=&-\frac{\partial \left[\bm{y}_s-\bm{\mu}(\bm{\rho})\right]^H\bm{C}^{-1}\left[\bm{y}_s-\bm{\mu}(\bm{\rho})\right]}{\partial \rho_{i}},\notag\\
=&-\frac{\partial \left[\bm{y}_s-\bm{\mu}(\bm{\rho})\right]^H}{\partial \rho_{i}}\bm{C}^{-1}\left[\bm{y}_s-\bm{\mu}(\bm{\rho})\right]\notag\\
&-\left[\bm{y}_s-\bm{\mu}(\bm{\rho})\right]^H\bm{C}^{-1}\frac{\partial \left[\bm{y}_s-\bm{\mu}(\bm{\rho})\right]}{\partial \rho_{i}},\notag\\
=&\,\frac{\partial\bm{\mu}(\bm{\rho})^H}{\partial\rho_{i}}\bm{C}^{-1}\left[\bm{y}_s-\bm{\mu}(\bm{\rho})\right]\notag
\\&+\left[\bm{y}_s-\bm{\mu}(\bm{\rho})\right]^H\bm{C}^{-1}\frac{\partial\bm{\mu}(\bm{\rho})}{\partial \rho_{i}},
\end{align}
and the second-order partial derivative of $\frac{\partial\,\small{\mbox{ln}}\bm{f}(\bm{y}_s|\bm{\rho})}{\partial \rho_{i}}$ with respect to $\rho_{j}$ is calculated as
\begin{align}
\frac{\partial^2\mbox{ln}\bm{f}(\bm{y}_s|\bm{\rho})}{\partial\rho_{j}\partial\rho_{i}}=&\frac{\partial}{\partial \rho_{j}}\bigg[\frac{\partial\bm{\mu}(\bm{\rho})^H}{\partial\rho_{i}}\bm{C}^{-1}\left[\bm{y}_s-\bm{\mu}(\bm{\rho})\right]\notag\\&+\left[\bm{y}_s-\bm{\mu}(\bm{\rho})\right]^H\bm{C}^{-1}\frac{\partial\bm{\mu}(\bm{\rho})}{\partial \rho_{i}}\bigg], 
\notag\\ 
\overset{(a)}{=}&\frac{\partial}{\partial \rho_{j}}\left[\frac{\partial\bm{\mu}(\bm{\rho})^H}{\partial\rho_{i}}\bm{C}^{-1}\left[\bm{y}_s-\bm{\mu}(\bm{\rho})\right]\right]\notag\\&+\frac{\partial}{\partial \rho_{j}}\left[\left[\bm{y}_s-\bm{\mu}(\bm{\rho})\right]^H\bm{C}^{-1}\frac{\partial\bm{\mu}(\bm{\rho})}{\partial \rho_{i}}\right],\notag\\
\overset{(b)}{=}&\frac{\partial^2\bm{\mu}(\bm{\rho})^H}{\partial\rho_{j}\partial\rho_{i}}\bm{C}^{-1}\left[\bm{y}_s-\bm{\mu}(\bm{\rho})\right]\notag\\-\frac{\partial\bm{\mu}(\bm{\rho})^H}{\partial\rho_{i}}&\bm{C}^{-1}\frac{\partial\bm{\mu}(\bm{\rho})}{\partial\rho_{j}}-\frac{\partial\bm{\mu}(\bm{\rho})^H}{\partial\rho_{j}}\bm{C}^{-1}\frac{\partial\bm{\mu}(\bm{\rho})}{\partial\rho_{i}}\notag\\&+\left[\bm{y}_s-\bm{\mu}(\bm{\rho})\right]^H\bm{C}^{-1}\frac{\partial^2\bm{\mu}(\bm{\rho})}{\partial\rho_{j}\partial\rho_{i}},
\end{align}
where equality (a) is obtained by the derivative sum rule and equality (b) is obtained by the derivative product rule. Therefore, the $(i,j)$-th element of FIM $\bm{J}_{\bm{\rho}}$, $\forall i,j\in[2L]$, is derived as  
\begin{align}
\bm{J}_{\bm{\rho}}[i,j]\overset{(a)}{=}&-\mathbb{E}\left\{\frac{\partial^2 \mbox{ln} \bm{f}(\bm{y}_s|\bm{\rho})}{\partial\rho_{j}\partial\rho_{i}}\right\},\notag\\
\overset{(b)}{=}&-\mathbb{E}\left\{\frac{\partial^2\bm{\mu}(\bm{\rho})^H}{\partial\rho_{j}\partial\rho_{i}}\bm{C}^{-1}\left[\bm{y}_s-\bm{\mu}(\bm{\rho})\right]\right\}\notag\\&-\mathbb{E}\left\{\left[\bm{y}_s-\bm{\mu}(\bm{\rho})\right]^H\bm{C}^{-1}\frac{\partial^2\bm{\mu}(\bm{\rho})}{\partial\rho_{j}\partial\rho_{i}}\right\}\notag\\
+\mathbb{E}&\left\{\frac{\partial\bm{\mu}(\bm{\rho})^H}{\partial\rho_{j}}\bm{C}^{-1}\frac{\partial\bm{\mu}(\bm{\rho})}{\partial\rho_{i}}+\frac{\partial\bm{\mu}(\bm{\rho})^H}{\partial\rho_{i}}\bm{C}^{-1}\frac{\partial\bm{\mu}(\bm{\rho})}{\partial\rho_{j}}\right\},\notag\\
\overset{(c)}{=}&\,2\mathbb{E}\left\{\mathcal{R}\left[\frac{\partial\bm{\mu}(\bm{\rho})^H}{\partial\rho_{i}}\bm{C}^{-1}\frac{\partial\bm{\mu}(\bm{\rho})}{\partial\rho_{j}}\right]\right\},\notag\\
\overset{(d)}{=}&\frac{2}{\sigma_{c}^2+\sigma_{z}^2}\, \mathbb{E}\left\{\mathcal{R}\left[\sum_{m=1}^{M}\frac{\partial\bm{\mu}[m,\bm{\rho}]^H}{\partial\rho_{i}}\frac{\partial\bm{\mu}[m,\bm{\rho}]}{\partial\rho_{j}}\right]\right\},
\end{align}
where equality (a) is obtained by the equality in \cite[Equation (3.11)]{Steven}, equality (b) is obtained by property of expectation, equality (c) is obtained due to $\bm{y}_s-\bm{\mu}(\bm{\rho})$ being a zero mean CSCG noise, and equality (d) is obtained by the definition of $\bm{\mu}(\bm{\rho})$ and $\bm{C}$ given in \eqref{Final}.

Then, the first-order partial derivatives of $\bm{\mu}[m,\bm{\rho}]$ with respect to $\tau_{0,l}$ and $\phi_{l}$, are respectively calculated as 
\begin{equation}
	\label{partialone}
	\frac{\partial\bm{\mu}[m,\bm{\rho}]}{\partial\tau_{0,l}}\overset{(a)}{=} \frac{\partial\bm{\mu}[m,\bm{\rho}]}{\partial n_{\tau_{l}}}\frac{\partial n_{\tau_{l}}}{\partial\tau_{0,l}}
	\overset{(b)}{=}\frac{s_l\alpha_l\bm{H}_{l}}{\Delta t }\frac{\partial\bm{x}[m-n_{\tau_{l}}]}{\partial n_{\tau_{l}}},
\end{equation}
and 
\begin{equation}
	\label{partialtwo}
	\frac{\partial\bm{\mu}[m,\bm{\rho}]}{\partial\phi_{l}}=s_l\alpha_l\dot{\bm{H}}_{l}\bm{x}[m-n_{\tau_{l}}],
\end{equation}
where equality (a) is obtained by the chain rule; equality (b) is obtained by the definition of $n_{\tau_{l}}$ given in \eqref{sample}; $\dot{\bm{H}}_l$ is the first-order partial derivative of $\bm{H}_{l}$ with respect to
$\phi_{l}$ and is calculated as $\dot{\bm{H}}_{l}=\frac{\partial \bm{a}_{r}(\phi_{l})}{\partial \phi_{l}}\bm{a}_{t}^T(\theta_l)=j2\pi\frac{d_R}{\lambda}\cos(\phi_{l})\bm{\Lambda}\bm{H}_{l}$, with $\bm{\Lambda}=\mbox{diag}\left\{0,1,\cdots,M_r-1\right\}$.

Next, to facilitate the derivation of \eqref{partone}-\eqref{partthree}, we present the following lemma.
\begin{Lemma}
\label{Key_Appendix}
	For matrix $\bm{H}_l$ defined in \eqref{sensing} and vector $\frac{\partial \bm{x}[m-n_{\tau_{l}}]}{\partial n_{\tau_{l}}}$ specified in \eqref{sample}, the following equality is invariably valid 
\begin{align}
	\label{Lemma_Appendix}
	&\mathbb{E}\left\{\mathcal{R}\left[\sum_{m=1}^{M}\left(\bm{H}_p\frac{\partial \bm{x}[m-n_{\tau_{p}}]}{\partial n_{\tau_{p}}}\right)^H\left(\bm{H}_q\frac{\partial \bm{x}[m-n_{\tau_{q}}]}{\partial n_{\tau_{q}}}\right)\right]\right\}\\
	\label{Lemma_Appendix_1}
	&=\begin{cases}
		\varepsilon_{g} \overline{F_g^2} N\Delta t \, \mbox{Tr} \left(\bm{H}_{p}\bm{R}_x\bm{H}_{p}^H\right),\;& p=q,\\
		0,\;& p\not=q.
	\end{cases}
\end{align}
\end{Lemma}

\indent \indent \emph{Proof:}
Based on the definition of the excitation signal, \eqref{Lemma_Appendix} is rewritten as
\begin{align}
\label{ThreeCases}
	\mathbb{E}\Bigg\{\mathcal{R}\bigg[\sum_{m=v_{p,q}+1}^{u_{p,q}+N}&\left(\bm{H}_p\frac{\partial \bm{x}[m-n_{\tau_{p}}]}{\partial n_{\tau_{p}}}\right)^H\notag\\
	&\left(\bm{H}_q\frac{\partial \bm{x}[m-n_{\tau_{q}}]}{\partial n_{\tau_{q}}}\right)\bigg]\Bigg\},
\end{align}
where $v_{p,q}=\max\left\{n_{\tau_{p}},n_{\tau_{q}}\right\}$ and $u_{p,q}=\min\left\{n_{\tau_{p}},n_{\tau_{q}}\right\}$. Then, based on the value of $p$ and $q$, \eqref{ThreeCases} is divided into three cases for comprehensive discussion.

\noindent \textbf{Case 1}: $p=q$. It is obviously that $n_{\tau_{p}}=n_{\tau_{q}}$, then \eqref{ThreeCases} is rewritten as 
	\begin{align}
		\label{ThreeCases-one}
		&\mathbb{E}\Bigg\{\sum_{m=n_{\tau_{p}}+1}^{n_{\tau_{p}}+N}  \left(\bm{H}_p\frac{\partial \bm{x}[m-n_{\tau_{p}}]}{\partial n_{\tau_{p}}}\right)^H\left(\bm{H}_p\frac{\partial \bm{x}[m-n_{\tau_{p}}]}{\partial n_{\tau_{p}}}\right)\Bigg\}\notag\\
		\overset{(a)}{=}&\, (\Delta t)^2\, \mathbb{E}\Bigg\{\sum_{m=n_{\tau_{p}}+1}^{n_{\tau_{p}}+N} \left(\bm{H}_{p}\dot{\bm{x}}(t)\right)^H\left(\bm{H}_{p}\dot{\bm{x}}(t)\right)\big\lvert_{t=(m-n_{\tau_{p}})\Delta t}\Bigg\}, \notag \\
		=&\, (\Delta t)^2\,\mathbb{E}\Bigg\{\sum_{n=1}^{N} \left(\bm{H}_{p}\dot{\bm{x}}(t)\right)^H\left(\bm{H}_{p}\dot{\bm{x}}(t)\right)\big\lvert_{t=n\Delta t}\Bigg\}, \notag \\
		=&\, \Delta t\,\mathbb{E}\left\{\int_{0}^{\Delta T} \left(\bm{H}_{p}\dot{\bm{x}}(t)\right)^H\left(\bm{H}_{p}\dot{\bm{x}}(t)\right) dt\right\}, \notag \\
		\overset{(b)}{=}& \Delta t\mathbb{E}\Bigg\{\sum_{n=1}^{N} \int_{(n-1)\Delta t}^{n \Delta t} \left(\bm{H}_{p}\bm{x}[n]\dot{g}(t')\right)^H\notag\\
		&\qquad\qquad\qquad\quad\left(\bm{H}_{p}\bm{x}[n]\dot{g}(t')\right)\big\lvert_{t'=t-(n-1)\Delta t}dt'\Bigg\}, \notag \\
		=&\, \Delta t\,\mathbb{E}\Bigg\{\sum_{n=1}^{N} \left(\bm{H}_{p}\bm{x}[n]\right)^H\left(\bm{H}_{p}\bm{x}[n]\right)\notag\\
		&\qquad\qquad\qquad\quad\int_{(n-1)\Delta t}^{n \Delta t} \big|\dot{g}(t')\big|_{t'=t-(n-1)\Delta t}^2 dt'\Bigg\}, \notag \\
		\overset{(c)}{=}& \varepsilon_{g} \overline{F_g^2} \Delta t\, \mathbb{E}\left\{\left(\mbox{vec} \left(\bm{D}_p\right)\right)^H\mbox{vec} \left(\bm{D}_p\right)\right\}, \notag \\
		\overset{(d)}{=}& \varepsilon_{g} \overline{F_g^2} \Delta t\,  \mathbb{E}\left\{\mbox{Tr}\left(\bm{D}_p^H\bm{D}_p\right)\right\}, \notag \\
		\overset{(e)}{=}&\varepsilon_{g} \overline{F_g^2} N\Delta t \, \mbox{Tr} \left(\bm{H}_{p}\bm{R}_x\bm{H}_{p}^H\right),
	\end{align}
	where $\dot{\bm{x}}(t)=\frac{\partial \bm{x}(t)}{\partial t}$ is the first-order partial derivative of $\bm{x}(t)$ with respect to $t$,  $\bm{D}_p=\bm{H}_p\bm{X}$ with $\bm{X}=\big[\bm{x}[1],\bm{x}[2],\cdots,\bm{x}[N]\big]$, equality (a) is obtained by the fact that 
	\begin{align}
		\frac{\partial \bm{x}[m-n_{\tau_{l}}]}{\partial n_{\tau_{l}}}&=\dot{\bm{x}}(t)\frac{\partial t}{\partial n_{\tau_p}}\bigg\lvert_{t=(m-n_{\tau_{p}})\Delta t}, \notag\\&=-\Delta t\dot{\bm{x}}(t)\big\lvert_{t=(m-n_{\tau_{p}})\Delta t},\;\;\forall p\in[L],
	\end{align}
	equality (b) is obtained by the definition of the excitation signal and transmit pulse function described in \eqref{transmit}, equality (c) is obtained by the fact that
	\begin{align}
		\int_{(n-1)\Delta t}^{n\Delta t} \big|\dot{g}(t')\big|_{t'=t-(n-1)\Delta t}^2 dt' &= \int_{0}^{\Delta t} |\dot{g}(t)|^2 dt, \notag \\&=\varepsilon_{g}\overline{F_g^2},
	\end{align}
	equality (d) is obtained by $\left(\mbox{vec}(\bm{A})\right)^H\mbox{vec}(\bm{A})=\mbox{Tr}(\bm{A}^H\bm{A})$, and equality (e) is obtained by $\mbox{Tr}(\bm{A}\bm{B})=\mbox{Tr}(\bm{B}\bm{A})$ and $\bm{X}\bm{X}^H=\sum_{n=1}^{N}\bm{x}[n]\bm{x}[n]^H=N\bm{R}_x$.
		
    \noindent \textbf{Case 2}: $p\not=q$ and $n_{\tau_{p}}<n_{\tau_{q}}$. \eqref{ThreeCases} is then rewritten as
	\begin{align}
		\label{ThreeCases-two}
		&\mathbb{E}\Bigg\{\mathcal{R}\bigg[\sum_{m=n_{\tau_{q}+1}}^{n_{\tau_{p}+N}}\left(\bm{H}_p\frac{\partial \bm{x}[m-n_{\tau_{p}}]}{\partial n_{\tau_{p}}}\right)^H\notag\\
		&\qquad\qquad\qquad\left(\bm{H}_q\frac{\partial \bm{x}[m-n_{\tau_{q}}]}{\partial n_{\tau_{q}}}\right)\bigg]\Bigg\} \notag\\
		=&\,(\Delta t)^2 \,\mathbb{E}\Bigg\{\mathcal{R} \bigg[\sum_{m=n_{\tau_{q}+1}}^{n_{\tau_{p}+N}}\left(\bm{H}_{p}\dot{\bm{x}}(t+\tau_{0,q})\right)^H\notag\\
		&\qquad\qquad \left(\bm{H}_{q}\dot{\bm{x}}(t+\tau_{0,p})\right)\big\lvert_{t=(m-n_{\tau_{p}}-n_{\tau_{q}})\Delta t}\bigg]\Bigg\},\notag\\
		=&\,(\Delta t)^2\, \mathbb{E}\Bigg\{\mathcal{R}\bigg[\sum_{n=1}^{N-\delta_{p,q}}\left(\bm{H}_{p}\dot{\bm{x}}(t+\delta_{p,q}\Delta t)\right)^H\notag\\
		&\qquad\qquad\qquad\qquad\qquad\left(\bm{H}_{q}\dot{\bm{x}}(t)\right)\big\lvert_{t=n\Delta t}\bigg]\Bigg\}, \notag \\
		=&\Delta t\mathbb{E}\Bigg\{\mathcal{R}\bigg[\int_{0}^{(N-\delta_{p,q})\Delta t}\left(\bm{H}_{p}\dot{\bm{x}}(t+\delta_{p,q}\right)^H\left(\bm{H}_{q}\dot{\bm{x}}(t)\right)dt \bigg]\Bigg\}, \notag \\
		=&\varepsilon_{g} \overline{F_g^2} \Delta t \, 
		\mathbb{E}\Bigg\{\mathcal{R}\bigg[\mbox{Tr}\big(\bm{H}_{q}\sum_{n=1}^{N-\delta_{p,q}}\bm{x}[n]\bm{x}[n+\delta_{p,q}]^H\bm{H}_{p}^H\big)\bigg]\Bigg\}, \notag \\
		\overset{(a)}{=}& \,0,
	\end{align}
	where $\delta_{p,q}=|n_{\tau_{p}}-n_{\tau_{q}}|$, and equality (a) is obtained under the assumption $\bm{x}[m]$ and $\bm{x}[n],\forall m\not=n$, are independent with each other, the cross-correlation matrix between them will be a zero matrix, i.e., 
	\begin{equation}
		\mathbb{E}\left\{\bm{x}[n]\bm{x}[m]^H\right\} =\bm{0},\quad \forall m\not=n.
	\end{equation}
	
	\noindent \textbf{Case 3}: $p\not=q$ and $n_{\tau_{p}}>n_{\tau_{q}}$. It is similar to the derivation of Case 2, \eqref{ThreeCases} is rewritten as
	\begin{align}
		\label{case-one-three}
		&(\Delta t)^2\,\mathbb{E}\Bigg\{\mathcal{R}\bigg[\sum_{n=1}^{N-\delta_{p,q}}\left(\bm{H}_{p}\dot{\bm{x}}(t)\right)^H\notag\\
		&\qquad\qquad\qquad\qquad\left(\bm{H}_{q}\dot{\bm{x}}(t+\delta_{p,q}\Delta t)\right)\big\lvert_{t=n\Delta t}\bigg]\Bigg\}, \notag\\
		=&\Delta t\mathbb{E}\Bigg\{\mathcal{R}\bigg[\int_{0}^{(N-\delta_{p,q})\Delta t}\left(\bm{H}_{p}\dot{\bm{x}}(t)\right)^H\notag\\
		&\qquad\qquad\qquad\qquad\left(\bm{H}_{q}\dot{\bm{x}}(t+\delta_{p,q}\Delta t)\right) dt \bigg]\Bigg\}, \notag \\
		=&\varepsilon_{g}\overline{F_g^2} \Delta t 
		\mathbb{E}\Bigg\{\mathcal{R}\bigg[\mbox{Tr}\left(\bm{H}_{q}\sum_{n=1}^{N-\delta_{p,q}}\bm{x}[n+\delta_{p,q}]\bm{x}[n]^H\bm{H}_{p}^H\right)\bigg]\Bigg\}, \notag \\
		=&\, 0.
	\end{align}

Based on the above analysis, Lemma \ref{Key_Appendix} has been proved.\QEDA

Finally, by substituting the first-order partial derivatives $\frac{\partial\bm{\mu}[m,\bm{\rho}]}{\partial\tau_{0,l}}$ and $\frac{\partial\bm{\mu}[m,\bm{\rho}]}{\partial\phi_{l}}$ into \eqref{Element}, and utilizing the Lemma \ref{Key_Appendix}, the $(p,q)$-th element of each component $\bm{G}_{\bm{\tau},\bm{\tau}}$, $\bm{G}_{\bm{\tau},\bm{\Phi}}$, and $\bm{G}_{\bm{\Phi},\bm{\Phi}}$ are derived. Specifically, the $(p,q)$-th element of the component $\bm{G}_{\bm{\tau},\bm{\tau}}$ is derived as
\begin{align}
\label{left-up}
&\bm{G}_{\bm{\tau},\bm{\tau}}[p,q]\notag\\
&=\frac{2}{\sigma_{c}^2+\sigma_{z}^2}\mathbb{E}\left\{\mathcal{R}\left[\sum_{m=1}^{M}\frac{\partial\bm{\mu}[m,\bm{\rho}]^H}{\partial\tau_{0,p}}\frac{\partial\bm{\mu}[m,\bm{\rho}]}{\partial\tau_{0,q}}\right]\right\},\notag\\
&=\frac{2s_ps_q\alpha_p\alpha_q}{(\sigma_{c}^2+\sigma_{z}^2)(\Delta t)^2}\times\notag\\
&\mathbb{E}\Bigg\{\mathcal{R}\bigg[\sum_{m=1}^{M}\left(\bm{H}_p\frac{\partial \bm{x}[m-n_{\tau_{p}}]}{\partial n_{\tau_{p}}}\right)^H\left(\bm{H}_q\frac{\partial \bm{x}[m-n_{\tau_{q}}]}{\partial n_{\tau_{q}}}\right)\bigg]\Bigg\},\notag\\
&=\begin{cases}
\frac{2\varrho\varepsilon_{g}\overline{F_g^2}Ns_p^2\eta(d_{T,p})\eta(d_{R,p})}{(\sigma_{c}^2+\sigma_{z}^2)\Delta t}\mbox{Tr}\left(\bm{H}_{p}\bm{R}_x(\bm{H}_{p})^H\right),&p=q,\\
0,&p\not=q,
\end{cases},
\end{align}
the $(p,q)$-th element of the component $\bm{G}_{\bm{\tau},\bm{\Phi}}$ is derived as
\begin{align}
\label{anti-diag}
&\bm{G}_{\bm{\tau},\bm{\Phi}}[p,q]\\
&=\frac{2}{\sigma_{c}^2+\sigma_{z}^2}\,\mathbb{E}\left\{\mathcal{R}\left[\sum_{m=1}^{M}\frac{\partial\bm{\mu}[m,\bm{\rho}]^H}{\partial\tau_{0,p}}\frac{\partial\bm{\mu}[m,\bm{\rho}]}{\partial\phi_{q}}\right]\right\},\notag\\
&=\frac{2s_ps_q\alpha_p\alpha_q}{(\sigma_{c}^2+\sigma_{z}^2)\Delta t}\times\\
&\quad\mathbb{E}\Bigg\{\mathcal{R}\bigg[\sum_{m=1}^{M}\left(\bm{H}_{p}\frac{\partial \bm{x}[m-n_{\tau_{p}}]}{\partial n_{\tau_{p}}}\right)^H\left(\dot{\bm{H}}_{q}\bm{x}[m-n_{\tau_{q}}]\right)\bigg]\Bigg\},\notag\\
&\overset{(a)}{=}\begin{cases}
\frac{2\varrho Ns_p^2\eta(d_{T,p})\eta(d_{R,p})}{(\sigma_{c}^2+\sigma_{z}^2)\Delta t}  \mathcal{R}\left[\varepsilon_{\dot{g}}\mbox{Tr}\left(\bm{H}_{p}\bm{R}_x\dot{\bm{H}}_{p}^H\right)\right],\;&p=q,\\
0,\;&p\not=q,
\end{cases},
\end{align}
and the $(p,q)$-th element of the component $\bm{G}_{\bm{\Phi},\bm{\Phi}}$ is derived as
\begin{align}
&\bm{G}_{\bm{\Phi},\bm{\Phi}}[p,q]=\\  
&=\frac{2}{\sigma_{c}^2+\sigma_{z}^2}\mathbb{E}\left\{\mathcal{R}\left[\sum_{m=1}^{M}\frac{\partial\bm{\mu}[m,\bm{\rho}]^H}{\partial\phi_{p}}\frac{\partial\bm{\mu}[m,\bm{\rho}]}{\partial\phi_{q}}\right]\right\},\notag \\ 
&=\frac{2s_ps_q\alpha_p\alpha_q}{\sigma_{c}^2+\sigma_{z}^2}\times\\
&\quad\;\mathbb{E}\left\{\mathcal{R}\left[\sum_{m=1}^{M}\left(\dot{\bm{H}}_{p}\bm{x}[n-n_{\tau_{p}}]\right)^H\left(\dot{\bm{H}}_{q}\bm{x}[n-n_{\tau_{q}}]\right)\right]\right\},\notag\\
&\overset{(b)}{=}\begin{cases} 
\frac{2\varrho \varepsilon_{g} Ns_p^2\eta(d_{T,p})\eta(d_{R,p})}{(\sigma_{c}^2+\sigma_{z}^2)\Delta t} \mbox{Tr}\left(\dot{\bm{H}}_{p}\bm{R}_x\dot{\bm{H}}_{p}^H\right),\;&p=q,\\
0 ,\;&p\not=q,
\end{cases},
\end{align}
where equality (a) is obtained by substituting $\bm{H}_{q}$ and $\frac{\partial\bm{x}[m-n_{\tau_{q}}]}{\partial n_{\tau_{q}}}$ with $\dot{\bm{H}}_{q}$ and $\bm{x}[m-n_{\tau_{q}}]$ in Lemma \ref{Key_Appendix}, respectively, and equality (b) is obtained by substituting $\bm{H}_{p}$, $\bm{H}_{q}$, $\frac{\partial\bm{x}[m-n_{\tau_{p}}]}{\partial n_{\tau_{p}}}$, and $\frac{\partial\bm{x}[m-n_{\tau_{q}}]}{\partial n_{\tau_{q}}}$ in Lemma \ref{Key_Appendix} with $\dot{\bm{H}}_{p}$, $\dot{\bm{H}}_{q}$, $\bm{x}[m-n_{\tau_{p}}]$, and $\bm{x}[m-n_{\tau_{q}}]$, respectively.

Based on the above analysis, Proposition \ref{Propositon_FIM} has been proved.

\section{Proof of Proposition \ref{CRBMatrix}}
\label{AppendixD}

\begin{figure*}
	\begin{equation}
		\label{Inverse}
		\bm{J}_{\bm{\rho}}^{-1}=\begin{bmatrix}
			\bm{G}_{\bm{\tau},\bm{\tau}}^{-1} + \bm{G}_{\bm{\tau},\bm{\tau}}^{-1}\bm{G}_{\bm{\tau},\bm{\Phi}}\bm{S}_{\bm{\tau},\bm{\tau}}^{-1}\bm{G}^T_{\bm{\tau},\bm{\Phi}}\bm{G}_{\bm{\tau},\bm{\tau}}^{-1} & -\bm{G}_{\bm{\tau},\bm{\tau}}^{-1}\bm{G}_{\bm{\tau},\bm{\Phi}}\bm{S}_{\bm{\tau},\bm{\tau}}^{-1} \\ 
			-\bm{S}_{\bm{\tau},\bm{\tau}}^{-1}\bm{G}^T_{\bm{\tau},\bm{\Phi}}\bm{G}_{\bm{\tau},\bm{\tau}}^{-1} & \bm{S}_{\bm{\tau},\bm{\tau}}^{-1} \\ 
		\end{bmatrix},  
	\end{equation}
	\hrulefill
\end{figure*}

In accordance with the Matrix Inversion Lemma \cite{Matrix}, the inverse of FIM $\bm{J}_{\bm{\rho}}$ given in \eqref{FIM} can be expressed as in \eqref{Inverse}, where $\bm{S}_{\bm{\Phi},\bm{\Phi}}$ and $\bm{S}_{\bm{\tau},\bm{\tau}}$ denote the Schur complement \cite{Schur} of $\bm{G}_{\bm{\Phi},\bm{\Phi}}$ and $\bm{G}_{\bm{\tau},\bm{\tau}}$, respectively. Moreover, since the matrices $\bm{G}_{\bm{\tau},\bm{\tau}}$, $\bm{G}_{\bm{\tau},\bm{\Phi}}$, and $\bm{G}_{\bm{\Phi},\bm{\Phi}}$ described in \eqref{partone}-\eqref{partthree} are diagonal matrices with identical dimensions, the Woodbury Formula \cite{Matrix}, i.e.,
\begin{equation}
	\bm{G}_{\bm{\tau},\bm{\tau}}^{-1} + \bm{G}_{\bm{\tau},\bm{\tau}}^{-1}\bm{G}_{\bm{\tau},\bm{\Phi}}\bm{S}_{\bm{\tau},\bm{\tau}}^{-1}\bm{G}^T_{\bm{\tau},\bm{\Phi}}\bm{G}_{\bm{\tau},\bm{\tau}}^{-1}= \bm{S}_{\bm{\Phi},\bm{\Phi}}^{-1},
\end{equation}
is employed to simplify \eqref{Inverse} into the following form
\begin{equation}
\label{Inverse_Element}
	\bm{J}_{\bm{\rho}}^{-1}=\begin{bmatrix}
\bm{S}_{\bm{\Phi},\bm{\Phi}}^{-1} & -\bm{G}_{\bm{\tau},\bm{\tau}}^{-1}\bm{G}_{\bm{\tau},\bm{\Phi}}\bm{S}_{\bm{\tau},\bm{\tau}}^{-1} \\ 
-\bm{S}_{\bm{\tau},\bm{\tau}}^{-1}\bm{G}^T_{\bm{\tau},\bm{\Phi}}\bm{G}_{\bm{\tau},\bm{\tau}}^{-1} & \bm{S}_{\bm{\tau},\bm{\tau}}^{-1} \\ 
\end{bmatrix},
\end{equation}
where the closed-form expressions of components $\bm{S}_{\bm{\Phi},\bm{\Phi}}$ and $\bm{S}_{\bm{\tau},\bm{\tau}}$ are respectively expressed as 
\begin{equation}
	\begin{split}
		\bm{S}_{\bm{\Phi},\bm{\Phi}}&=\bm{G}_{\bm{\tau},\bm{\tau}}-\bm{G}_{\bm{\tau},\bm{\Phi}}\bm{G}_{\bm{\Phi},\bm{\Phi}}^{-1}\bm{G}_{\bm{\tau},\bm{\Phi}}^T,\\
		&= \varepsilon_{g}\overline{F_g^2}\,\bm{\xi}\cdot\bm{B}_{\bm{\tau},\bm{\tau}}-\frac{1}{\varepsilon_{g}}\bm{\xi}\cdot\bm{B}_{\bm{\tau},\bm{\Phi}}^2\cdot\bm{B}_{\bm{\Phi},\bm{\Phi}}^{-1},
	\end{split}
\end{equation} 
and 
\begin{equation}
	\begin{split}
		\bm{S}_{\bm{\tau},\bm{\tau}}&=\bm{G}_{\bm{\Phi},\bm{\Phi}}-\bm{G}_{\bm{\tau},\bm{\Phi}}^T\bm{G}_{\bm{\tau},\bm{\tau}}^{-1}\bm{G}_{\bm{\tau},\bm{\Phi}},\\
		&= \varepsilon_{g}\,\bm{\xi}\cdot\bm{B}_{\bm{\Phi},\bm{\Phi}}-\frac{1}{\varepsilon_{g}\overline{F_g^2}}\bm{\xi}\cdot\bm{B}_{\bm{\tau},\bm{\Phi}}^2\cdot\bm{B}_{\bm{\tau},\bm{\tau}}^{-1}.
	\end{split}
\end{equation}
Therefore, the closed-form expressions for the CRB on transmission delay $\bm{\tau}$ and DoA $\bm{\Phi}$ are respectively derived as 
\begin{align}
	\mbox{CRB}(\bm{\tau})&=\mbox{Tr}\left(\bm{S}_{\bm{\Phi},\bm{\Phi}}^{-1}\right), \notag \\
	&=\sum_{l=1}^{L} \frac{1}{\varepsilon_{g}\overline{F_g^2}\xi_l\bm{B}_{\bm{\tau},\bm{\tau}}[l,l]-\frac{\xi_l}{\varepsilon_{g}}\frac{|\bm{B}_{\bm{\tau},\bm{\Phi}}[l,l]|^2}{\bm{B}_{\bm{\Phi},\bm{\Phi}}[l,l]}} , \notag \\
	&=\sum_{l=1}^{L} \frac{\varepsilon_{g}h_l(\bm{R}_x)}{\varepsilon_{g}^2\overline{F_g^2}f_l(\bm{R}_x)h_l(\bm{R}_x)-|\varepsilon_{\dot{g}}|^2g_l^2(\bm{R}_x)}\times\frac{1}{\xi_l},
\end{align}
and
\begin{align}
	\mbox{CRB}(\bm{\Phi})&=\mbox{Tr}\left(\bm{S}_{\bm{\tau},\bm{\tau}}^{-1}\right), \notag \\
	&=\sum_{l=1}^{L} \frac{1}{\varepsilon_{g}\xi_l\bm{B}_{\bm{\Phi},\bm{\Phi}}[l,l]-\frac{\xi_l}{\varepsilon_{g}\overline{F_g^2}}\frac{|\bm{B}_{\bm{\tau},\bm{\Phi}}[l,l]|^2}{\bm{B}_{\bm{\tau},\bm{\tau}}[l,l]}}, \notag \\
	&=\sum_{l=1}^{L}\frac{\varepsilon_{g}\overline{F_g^2}f_l(\bm{R}_x)}{\varepsilon_{g}^2\overline{F_g^2}h_l(\bm{R}_x)f_l(\bm{R}_x)-|\varepsilon_{\dot{g}}|^2g_l^2(\bm{R}_x)}\times\frac{c_l}{\xi_l}. 
\end{align}
Based on the above analysis, Proposition \ref{CRBMatrix} has been proved.

\end{appendices}

\bibliographystyle{IEEEtran}   
\bibliography{isac}

\end{document}